\documentclass[a4paper,twoside]{ar-1col}
\usepackage[numbers]{natbib}

\usepackage[hidelinks]{hyperref}
\usepackage{
  amsmath,
  soul,
  ragged2e
}
\usepackage[marginparwidth=4.5cm, marginparsep=3mm]{geometry}

\jname{Xxxx. Xxx. Xxx. Xxx.}
\jvol{AA}
\jyear{YYYY}
\doi{10.1146/((please add article doi))}

\begin{document}

\markboth{Peinke Tabar Waechter}{Fokker-Planck Approach}

\title{The Fokker-Planck Approach to Complex Spatio-Temporal Disordered Systems}

\author{J. Peinke,$^1$. M. R. R. Tabar,$^2$  and M. W\"achter$^1$
\affil{$^1$
Institute of Physics and ForWind, University of Oldenburg, Oldenburg, D - 26111, e-mail: joachim.peinke@uni-oldenburg.de}
\affil{$^2$Dept. of Physics, Sharif University of Technology,Theran, Iran}
}

\begin{abstract}
When the complete understanding of a complex system is not available, as, e.g., for systems considered in the real-world, we need a top-down approach to complexity. In this approach one may start with the desire to understand general multi-point statistics. Here such a general approach is presented and discussed based on examples from
turbulence and sea waves. Our main idea is based on the cascade picture of turbulence, entangling fluctuations from large to small scales. Inspired by this cascade picture, we express the general multi-point 
statistics by the statistics of scale-dependent fluctuations of variables and relate it to a scale-dependent  process, which finally is a stochastic cascade process. We show how to extract from empirical data a Fokker-Planck equation for this cascade process, which
allows to generate surrogate data to forecast extreme events as well as to develop
a non-equilibrium thermodynamics for the complex systems. For each cascade events an
entropy production can be determined. These entropies fulfil accurately
a rigorous law, namely the integral fluctuations theorem.
\end{abstract}

\begin{keywords}
  multi-point statistics, stochastic process, Fokker-Planck equation,
  self-similarity, short time forecast, non-equilibrium
  thermodynamics, integral fluctuation theorem
\end{keywords}
\maketitle

\tableofcontents



\section{INTRODUCTION}
\label{Sec:Intro}

For quite some time research on complex systems has been considered as a continuation of investigating nonlinear or chaotic dynamics. The main difference between these systems may be understood by realizing that nonlinear or chaotic systems are spatially homogeneous and, thus, are described by low-dimensional nonlinear differential equations, cf. \cite{Argyris2015,heslot1987transitions}, whereas complex systems possess spatial and temporal inhomogeneities. Due to the interdependence, relationships, or interactions between units of a complex system the understanding of the entire system is not attainable by simply understanding each part, or by the local features. Complex systems are in general composed of many interacting subunits, where nonlinearities play an important role, so that complex spatio-temporal structures emerge; see e.g. \cite{Haken1,bar1997dynamics}. A consequence of the interaction between the subsystems and the overall behaviour is that it is often difficult to achieve full comprehensive understanding of complex systems dynamics. Additionally, surprising new collective phenomena may emerge. Examples of emergent behaviours include short- and long-term climate changes, hurricanes,
cascading failures, evolution, learning, and intelligence, to name just a few \cite{Friedrich2011}.

\begin{figure}[htbp]
\begin{center}
 \includegraphics[width=3in]{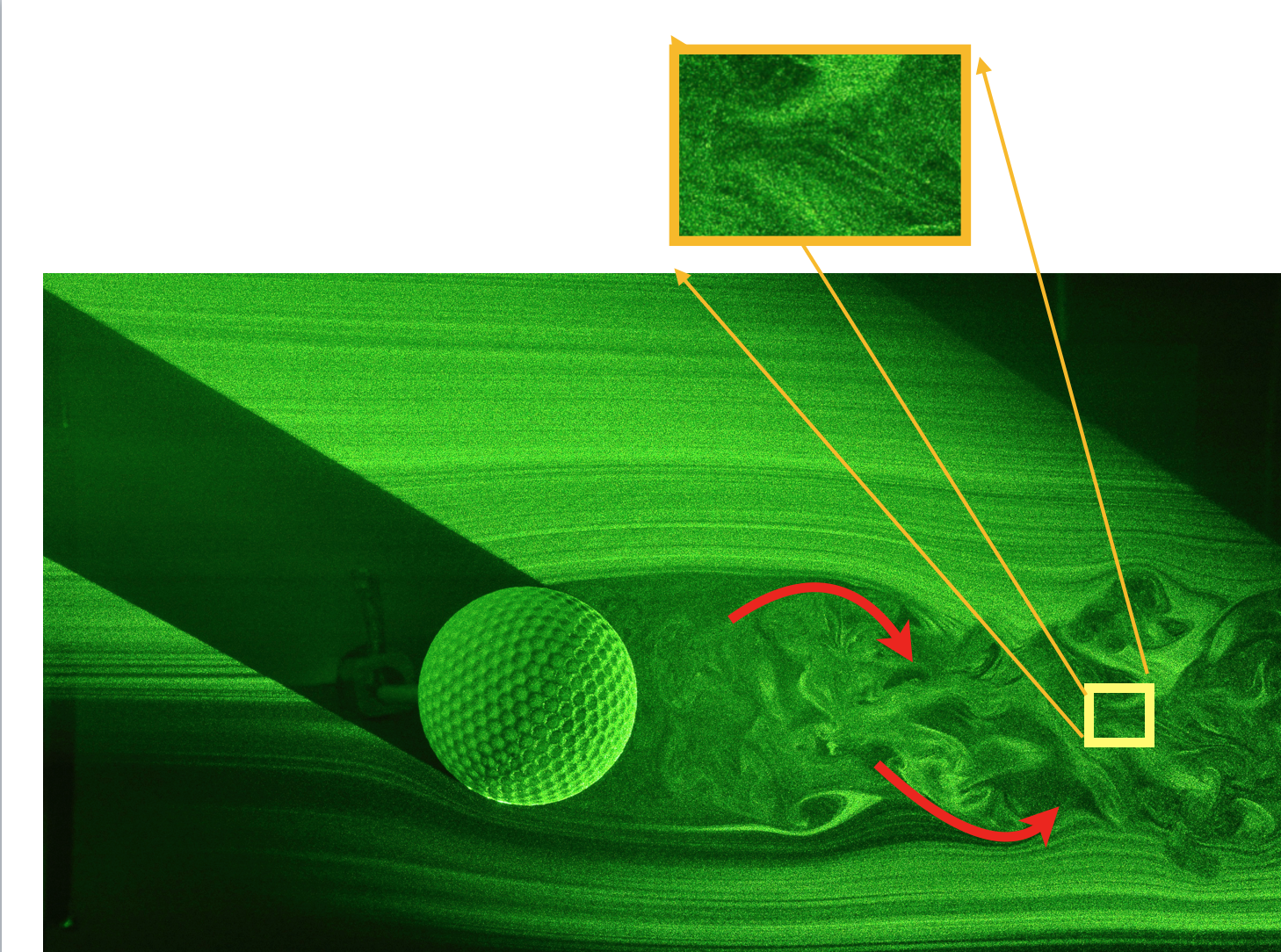}
\caption{Smoke visualisation of flow patterns of a turbulent wake
  behind a sphere with dimples. The overall structure of the wake flow
  is recognised right away by its large structures, highlighted by
  arrows. A closer look at this turbulent wake will show that
  flow patterns are never repeated exactly either in time or space,
  see blow-up. For many complex disordered systems the challenge is to
  understand this interplay between clear structures and
  stochasticity. }
\label{Fig_turb_flow}
\end{center}
\end{figure}

In this article we will take turbulent flows and sea waves as examples of complex systems. The main task for a good understanding of
the appealing complexity of flow patterns, like shown in Fig.~\ref{Fig_turb_flow}, is to characterise the clearly visible structures, as well as their large variability \cite{Frisch2001}. Watching such turbulent flows, one recognises immediately the flow type by its overall structure, but at the same time one gets the impression that over the time exactly same patterns are never seen twice. This mutuality of order and stochasticity is one exciting aspect of flow patterns. The two examples, selected for this article, reflect also nicely this mutuality. One of the challenging problems of turbulence is the small scale structure and its deviation from Gaussian statistics, cf. \cite{Frisch2001,Davidson2004,Pope2000}. The anomalous statistics can be seen in connection with the millennium problem, defined by Clay Mathematic Institute, where it is asked for the local structure of a solution of turbulent flows described by the Navier-Stokes equation \cite{Clay}. An open question is whether there are special small scale coherent structures explaining the anomalous statistics. For sea waves coherent structures seem natural, but here we should note that we are not interested in the case of periodic wave structures but in the cases of the rough sea. Hereof a most prominent wave structure is the monster wave, also called freak or rogue wave. Still an open question remains whether these structures are part of the disordered wave state or somehow independent of it (cf. \cite{Nazarenko2016,Onorato2013,Akhmediev2009}). Turbulent-like features of waves are treated as wave turbulence.
This leads to the open problem of what the basic features of such complex structures are. The questions can be formulated, whether there are some clear structures (coherent deterministic structures) that can be singled out and may serve as a skeleton to access the complexity.  Alternatively, one may ask, whether such systems can be understood best by their stochasticity and statistics? 
Quite often one approaches such complex systems in a pragmatic way by either studying the structures or the stochasticity.

Another common approach to characterise such complex patterns is to apply linear correlation measures, as such patterns decay in time or, respectively, in space.  Typically for complex systems, like for turbulence, no simple exponential decay of such correlations is found, consequently we are facing the problem of multi-scale correlated systems. Indeed, for such systems the low order statistics, i.e. one point statistics and two-point correlations, are not sufficient to grasp the observed complexity. 

Our present work has to be seen in a clear difference to the works devoted on the analysis of nonlinear dynamics with and without noise in the framework of time series analysis.  For time series anaylsis the proper embedding, Lyapunov-exponents, fractal dimensions,  fixed points, stable and unstable limit cycles, reconstruction of dynamic equations etc. are of interest, cf. \cite{Kantz2004,Friedrich2011}. Although methods of time series analysis will be used, this work focuses on the extended spatial disordered structures with the aim to get a comprehensive characterisation at arbitrarily many points, i.e. by a general N-point joint statistics. The development of such a method has been stimulated by research on turbulent flows \cite{Friedrich1997} and may be considered as a top-down approach.  Knowing such a general description, it should be possible to determine all statistical aspects of the system. Moreover, the common problem of structures versus statistics should be sorted out, as the general joint N-point statistics can grasp any sequential ordering of some patterns as multi-point structures. Also the mentioned multi-scale correlations and higher order statistics can be captured by general N-point statistics. The question is how complicated such joint statistics become. For the empirical estimation of N-point statistics the question whether sufficient data can be provided rises immediately.

For a general approach to N-point statistics, we propose a hierarchical ordering of the N-point statistics. 
This hierarchical ordering is in analogy to the common cascade picture of turbulence, which describes how structures on larger scales interact with structures on smaller scales in a hierarchical way, so that a downwards cascade from large to small scales is obtained.
Inspired by the idea of a cascade we investigate how the scale-dependent structures will change with the scale at each location. Vividly interpreted, this can be taken as a zooming-in process of the complex structure. To get access to the high demanding multi-point statistics, we set the scale-dependency of the complex structure in the context of a stochastic process evolving in scale. The novelty is that we do not consider  the common time-evolution of stochastic processes but an evolution in scale. In particular we show evidence that this zooming-in process can be approximated by a Markov process, i.e. that this process has no memory in its scale evolution. This approximation allows to derive a Fokker-Planck equation, evolving in scale, for the hierarchical ordering of the N-point statistics. The Fokker-Planck equation is not only a compact description of the whole complexity, but also enables to derive several other aspects ranging from scaling behaviour to thermodynamics, as outlined in this paper.

This work has its origin in a series of former works started in 1996 \cite{Peinke1996,Friedrich1997,Friedrich1997b}.  Initially the idea of a scale-dependent process has been worked out without paying much attention to how to nest  larger structures into smaller structures, and can be seen as a continuous formulation of the propagator description of the cascade \cite{amblard1999cascade,Castaing1990}.
Reviews on this approach can be found in \cite{Friedrich2011,Friedrich2012}, where the stochastic processes in general and the difference between commonly known stochastic processes in time and the new processes in scale are worked out.  In \cite{Friedrich2011} applications and citations of stochastic processes in scale are given, which range from turbulent flows, financial data and surface roughness to earthquakes, cosmic background radiation and iEEG recordings from epilepsy patients. All these examples posse remarkable multi-scale features and the complexity seems to be related to a hierarchical ordering connected to cascade-like structures. At this stage, the expression of multi-point and multi-scale statistics was used more or less synonymously.
With the attempt to reconstruct time series from the knowledge of the multi-scale processes \cite{Nawroth2006} the meaning of the correct placement of the smaller scale structures within the larger ones became clear and the relation between mutli-scale and multi-point statistics has been worked out. In \cite{Nawroth2010} this was done for financial market data and a short-time forecasting 
has been worked out. For turbulence data \cite{Stresing2010} it has been realised that  an extended class of stochastic cascade processes, expressed by a family of Fokker-Planck equations, is needed. In the present work we work out in detail the multi-point approach and relate this to the stochastic cascade processes in scale. The technical details for the handling of empirical data will be given in the corresponding sections. 

In our approach, we first consider the simplification of one-dimensional cuts of complex patterns (for turbulence this simplification is related to the Taylor's hypothesis of frozen turbulence). Thus, a quantity $q(x)$ along an axis $x$ is considered. Second, the hierarchical ordering is introduced by changing the scales $r$, thus we ask how the structure looks like on different scales $r$, where the changes go from large to small distances, as explained in the next section. Furthermore we show how the N-point statistics can be expressed by a joint multi-scale statistics. In Sec. 3  a three-point approach or three-point closure for the hierarchical multi-scale statistics is described, which finally opens up the possibility of projecting the general N-point statistics on the stochastic processes in the scale parameter $r$, ending in a scale-dependent Fokker-Planck equation, see Sec. 4. In Sec. 5 special self-similar or fractal solutions of the stochastic cascade process are discussed. Two further consequences are deduced from this approach. On the one side, we show in Sec. 6 that surrogate data sets can be produced with the same statistical properties and patterns (original processes). On the other side, in Sec. 7, the stochastic approach is put in the context of non-equilibrium thermodynamics for the complex systems, relating complex structures with the general fluctuation theorem.


\begin{textbox}[h]
\section{TAYLOR'S HYPOTHESIS OF FROZEN TURBULENCE}

In 1938 G. I. Taylor introduced a hypothesis by which he deduced the spatial fluctuations of a turbulent velocity profile 
from the corresponding measurements of temporal fluctuations at a single point. This hypothesis, known as the Taylor frozen-flow hypothesis, relies on the existence of a
mean flow $ \langle u \rangle $ that translates the spatial structures past a stationary probe in a time smaller than the inherent evolution time of the fluctuations \cite{Taylor1938}.  $q(x+dx) \approx q(x- \langle u \rangle dt)$, where we take the mean flow in $x$-direction.
Therefore, analyses are taken to be equivalent regardless of whether they are taken as snapshot in space (see Fig.~\ref{Fig_turb_flow}) or as time sequence of the structures passing over a sensor by the mean flow velocity. Here, we discuss only the spatial complexity.

\end{textbox}



\section{MULTI-POINT STATISTICS EXPRESSED BY INCREMENT STATISTICS }
\label{Sec:multi-point}

One important basic aspect of this work is the connection between the general multi-point characterisation of a complex structure and its multi-scales properties. We start with a quite formal consideration and show how  this connection can be worked out mathematically, the next sections show consequences and apllications. We consider the case that a complex structure is given as space and time dependence of a quantity $\vec{q}(\vec{x},t)$. For the example of a turbulent flow  $\vec{q}(\vec{x},t)$ is given by a velocity field $\vec{u}(\vec{x},t)$. For a surface $\vec{q}(\vec{x},t)$  is the spatial pattern of the height $h(\vec{x},t)$. 
As mentioned in the introduction, we simplify this system by assuming that its temporal and spatial structures are statistically similar, and in addition only one direction is of interest. We also assume that the characterising quantity is a scalar $q(x)$. We are interested in multi-point statistics, i.e. the probability of finding a sequence of events $q(x_i)$ for several discrete locations $x_i$ with $i=0,\cdots,N$, which is given by the joint probability density function (jPDF)
\begin{equation}
W(q_0,q_1,\cdots,q_N) \quad
\label{joint_pdf}
\end{equation}
where we used the abbreviation $q_i := q(x_i)$. Here $W(q_0,q_1,\cdots,q_N) dq_0  \cdots dq_N $ is the probability that the random variables $q(x_0),q(x_1), \cdots, q(x_N)$ belong to the intervals  $q_0\leq q(x_0) \leq q_0+dq_0, \cdots,  q_N\leq q(x_N) \leq q_N+dq_N$.  Instead of this joint $N+1$-point PDF $W$, one may be interested in the conditional probability of obtaining the value q at one selected point under the condition of the remaining events. 

In the following we select the value of $q$ at the last point $x_N$ as reference value.
Therefore the conditional probability of finding  $q(x_N)$ for the given preceding data $q(x_i)$, with $i=0,\ldots,N-1$, is given by the conditional probability density function (cPDF), which can also be taken as a transition probability, 
\begin{equation}
p(q_N| q_0,\ldots,q_{N-1})=\frac{W(q_0,q_1,\ldots,q_N)}{W(q_0,\ldots,q_{N-1})} .
\label{eq:c_pdf}
\end{equation}

\begin{figure}[htbp]
  \begin{center}
    \includegraphics[width=3.5in]{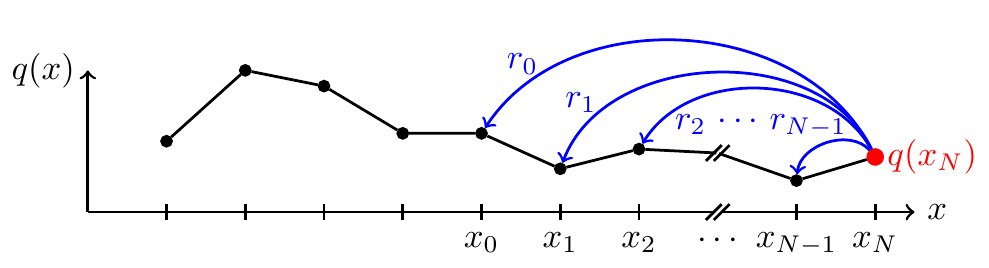}
    \caption{Scheme of the hierarchical ordering of increments
      $\xi_i = q(x_N) - q(x_N - r_i)$ and scales $r_i$, in order to
      describe $N$-point statistics. Sorting the scales after their
      sizes, from large to small, one obtains a zooming-in process, to
      which a trajectory $\xi_i(\cdot)$ can be assigned.
      This process then decribes the evolution of the increments
      $\xi_i$ when scales $r_i$ evolve from largest to smallest size.}
  \label{const_inc}
\end{center}
\end{figure}

The multi-point probabilities can also be expressed in another way by considering the statistics of relative changes from one selected point. For $x_N$ as the point of reference we denote the distance $r_i := x_N - x_i$. Therefore we introduce increments (other notations are common in the literature like $\delta_r q(x_i), \Delta q(x_i), q_r(x_i), ...$)
\begin{equation}
\xi_i := \xi (x_N,r_i) = q(x_N) - q(x_N - r_i)  
\label{inc}
\end{equation}
for  $i = 0,\ldots, N-1$, which quantify the differences of $q$ over the distances or scales $r_i$, as illustrated in Fig. \ref{const_inc}.

Using the coordinate transformation $q_i = q_N - \xi_i$, the $N+1$-point jPDF of Equation (\ref{joint_pdf}) can be
rewritten without loss of information as a jPDF of $N$ increments and the reference value $q_N$
\begin{eqnarray}
W(q_0,\ldots,  q_N)~  dq_0\ldots dq_N &=& 
W(\xi_0,\xi_2,\ldots,\xi_{N-1},q_N) ~ |J| ~ d\xi_0 d\xi_2\ldots d\xi_{N-1}dq_N \cr \nonumber \\
&=&p(\xi_0,\xi_2,\ldots,\xi_{N-1}|q_N)\cdot W(q_N) ~ d\xi_0 d\xi_2\ldots d\xi_{N-1}dq_N  .
\label{eq:Winc}
\end{eqnarray}
 $|J|$ denotes the determinate of Jacobin for transformation $(q_0,\ldots,  q_N) \to \xi_0,\xi_2,\ldots,\xi_{N-1},q_N$ and is unity. 
Here $W(q_N)$ is the one-point probability density function (PDF) for the value
$q_N$. Based on the natural ordering $x_0 < x_1 <\ldots <x_N$ the scales
$r_i$ are ordered as $r_i > r_{i+1}$. Note that  we have defined a scale evolution of
$r_i$ running with the index $i$ from large to small scale, which can be illustrated as a  process where one zooms in to resolve
smaller and smaller structures.

With Equation (\ref{eq:Winc}) we have expressed the
general $N+1$-point statistics by the statistics of $N$ increments
$\xi_i$ taken in a right-justified way from the point $q_N$. If the
statistics of the complex structure is homogeneous (or, for time
dependencies, stationary) the probability
$W(\xi_0,\xi_2,\ldots,\xi_{N-1},q_N) $ does not depend on the location
$x_N$ but  on the values of the describing quantity, here $q_N$.

\begin{marginnote}[]
\entry{describing quantity}{$q(x_i) = q_i$}
\entry{relative distance}{$r_i = x_N - x_i$}
\entry{increment}{$\xi_i := \xi (x_N,r_i) = q(x_N) - q(x_N - r_i)  $ there are other common notations like $q_r$ $\delta q$.}
\end{marginnote}

\begin{textbox}[h]
\section{MULTI-POINT AND MULTI-SCALE  STATISTICS}


A clear differentiation between  multi-point statistics and multi-scale  statistics should be made. 
The {\bf multi-point statistics} is given by $W(q_0,\ldots,  q_N)$. After equation \ref{eq:Winc} the multi-point statistics can also
be expressed in terms of   increments $\xi_i$  
$$W(q_0,\ldots,  q_N)\simeq W(\xi_0,\xi_1,\ldots,\xi_{N-1},q_N) ,$$
The {\bf multi-scale  statistics} is 
$$ W(\tilde{\xi}_0,\tilde{\xi}_1,\ldots,\tilde{\xi}_{N-1})$$
for the scale-dependent quantity $\tilde{\xi}_i (r_i)$, which could be besides an increment also another scale-dependent quantity (see Sidebars on Wavelets). 

  When increments $\xi_i=\xi(x_N,r_i)$ are considered for multi-point  statistics, 
  the definition of the reference point $x_N$ is of special
  relevance. Besides the  \emph{left-jusitified} increment
  $\xi^l_i(x_N,r_i)=q(x_N+r_i)-q(x_N)$ also a \emph{right-justified}
  definition $\xi^r_i(x_N,r_i)=q(x_N)-q(x_N-r_i)$ can be used, as we do here. For the multi-scale statistics
also a centered version $\xi^c_i(x_N,r_i)=q(x_N+r_i/2)-q(x_N-r_i/2)$ has been used,
  cf.~\cite{Waechter2004b, Waechter2004}, where further details on the relation between these different definitions are given. 
In this paper we restrict ourselves to the right-justified definition, as this is required for the multi-point reconstruction of data in section
 \ref{Sec:sorrogate}.  
 
 Note that from the multi-point statistics the multi-scale statistics can be derived, but as the reference value $q_N$ is not any more taken explicitly into account in the multi-scale statistics, one can not derive the multi-point statistics from the multi-scale statistics in general, see also \cite{Nawroth2010,Stresing2010}.

\end{textbox}



\section{CLOSURES OF MULTI-POINT STATISTICS}
\label{Sec:closure}

The introduction of the hierarchical ordering  of the increments $\xi_0,\xi_1,\ldots,\xi_{N-1}$, with $r_i > r_{i+1}$, (see Fig.~\ref{const_inc}) has been used so far just to reformulate the N-point statistics by increment statistics. As a next step we consider  this hierarchical ordering  of the increments as  scale-dependent fluctuations of the quantity $q$ that go from large to small scales or vice versa.  Later on, this will become an essential aspect for working out a cascade idea for the description of complex systems. Before we come to this point some formal aspects of the joint probabilities have to be discussed. The jPDF of Equation \ref{eq:Winc} can be expressed by a product of conditional probabilities
\begin{eqnarray} \label{eq:Winc_cond}
  W(\xi_0,\xi_2,\ldots,\xi_{N-1},q_N) =
  p(\xi_{N-1}|\xi_{N-2},\ldots,\xi_0,q_N) \cdot
  p(\xi_{N-2}|\xi_{N-3},\ldots,\xi_0,q_N) \cdot\ldots\cdot
  p(\xi_{1}|\xi_0,q_N)\cdot p(\xi_{0}|q_N) \cdot W(q_N) .
  \label{eq:Winc2}
\end{eqnarray}
A tremendous simplification arises, if the multi-conditioned PDF only depends on the increment of the next larger scale 
\begin{equation}
p(\xi_i|\xi_{i-1},...,\xi_0,q_N) = p(\xi_i|\xi_{i-1},q_N)  .
\label{eq:markov}
\end{equation}
We then obtain a much simpler form of Eq. (\ref{eq:Winc_cond}), i.e.
\begin{equation}
  W(\xi_0,\xi_1,\ldots,\xi_{N-1},q_N) =
  p(\xi_{N-1}|\xi_{N-2},q_N)\cdot  p(\xi_{N-2}|\xi_{N-3},q_N)
  \cdot\ldots\cdot
  p(\xi_{0}|q_N) \cdot W(q_N) .
\label{eq:Winc_mar}
\end{equation}
Note that these simplified cPDF $p(\xi_i|\xi_{i-1},q_N)$ are three-point
statistics, $p(\xi_i|\xi_{i-1},q_N)\cdot W(\xi_{i-1},q_N)  \simeq W(q_{i-1}, q_i, q_N) $.  Therefore Equation \ref{eq:Winc_mar} is a three-point
closure of the the general $(N+1)$-point jPDF.  As a remark we  mention, that possible closures are of central interest for the turbulence problem, and that in \cite{FriedrichJDiss} such a three point closure is discussed for the Lundgren-Monin-Novikov Hierarchy, a description of turbulence by multi-point probabilities.

Two further simplifications are given, first, if the cPDF are independent of the reference $q_N$
\begin{equation}
p(\xi_i|\xi_{i-1},q_N) = p(\xi_i|\xi_{i-1})  
\label{eq:ind_qN}
\end{equation}
 and second, if the cPDF are independent of larger increments
\begin{equation}
p(\xi_i|\xi_{i-1},q_N) =  p(\xi_i | q_N)  \;   \text{or}  \; p(\xi_i|\xi_{i-1}) = W(\xi_i) .
\label{eq:two_point}
\end{equation}
 The last conditions correspond to a two-point closure for which the general
 $N+1$-point probability factorises completely to products of simple one increment or one scale probabilities as
\begin{eqnarray}
W(\xi_0,\xi_1,\ldots,\xi_{N-1},q_N) &=&
p(\xi_0 | q_N)\cdot  p(\xi_1 | q_N) \cdot\ldots\cdot p(\xi_{N-1}|q_N)  W(q_N)\\
W(\xi_0,\xi_1,\ldots,\xi_{N-1}) &=&
W(\xi_0)\cdot  W(\xi_1) \cdot\ldots\cdot W(\xi_{N-1}) ,
\label{ninc_mar_2pt}
\end{eqnarray}
rely on the dependence on $q_N$. Only for the case that the increment statistics are also independent of the reference value $p(\xi_i|q_N)=W(\xi_i)$, a complete knowledge of the single increment PDFs $W(\xi_i) $ for all scales characterise completely the multi-scale disorder of the considered complex structure, an aspect which is important for the fractal characterisation of complex structures, see also Sec. \ref{Sec:fractal}. We can conclude that a characterisation of a
complex system, which is done only by the statistics of increments based on $W(\xi_i)$, is a special two-point characterisation.
With the knowledge of the PDF  $W(\xi_i)$ all higher order moments of $\langle\xi_i^n\rangle$ for all scales are known, but nothing
is known on more than two-point correlations.
Thus the question of how far the general multi-point problem can be
reduced is of central importance for  the proper characterisation of complex
systems.

\begin{figure}
  %
  %
  \includegraphics[width=\linewidth]{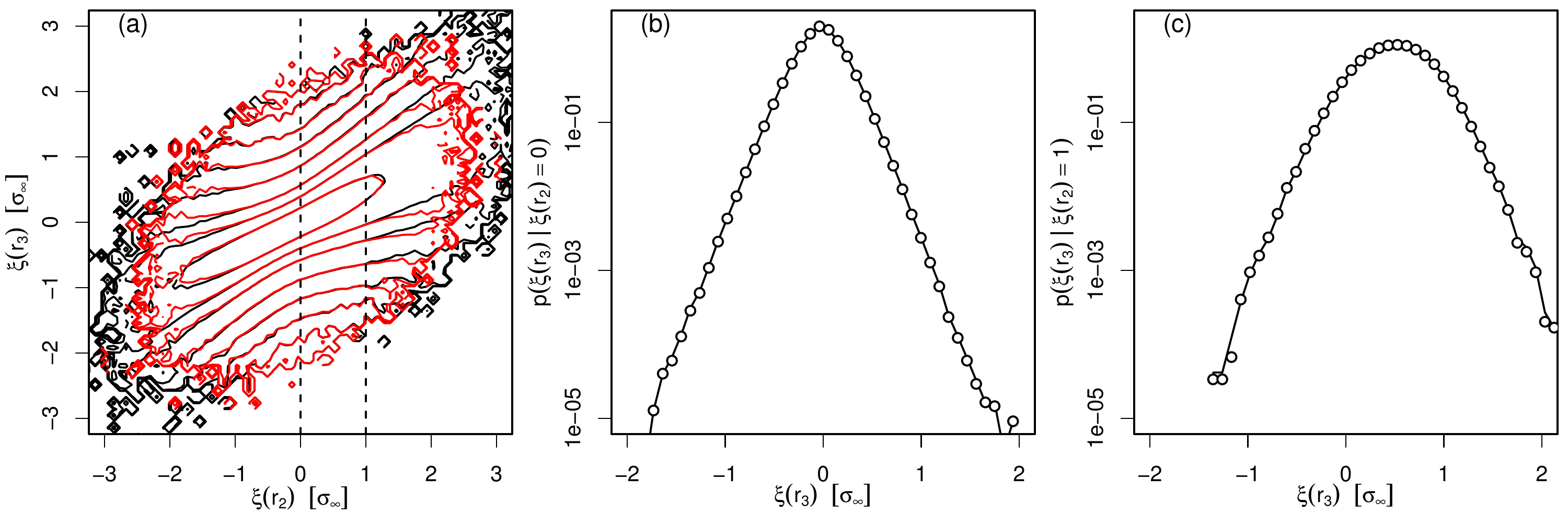}
  \caption{Visualization of Markov properties for velocity increments
    $\xi(r)$ in a turbulent flow with a Taylor microscale of
    $\lambda=0.0024$\,m after \cite{Stresing2010}. Here, the scales $r_i$ of the increments are
    $r_1=3\lambda$, $r_2=2\lambda$, and $r_3=\lambda$.
    (a) Contours of single- and double-conditional PDF
    $p(\xi(r_3)\,|\,\xi(r_2))$ (black) and $p(\xi(r_3)\,|\,\xi(r_2),
    \xi(r_1)=0)$ (red). Cuts at $\xi(r_2)=0$ and $\xi(r_2)=1$ are
    marked by vertical dashed lines.
    (b, c) Cuts of (a) at $\xi(r_2)=0$ and $\xi(r_2)=1$.
    Single-conditional PDF $p(\xi(r_3)\,|\,\xi(r_2))$ are presented as
    solid lines, double-conditional PDF $p(\xi(r_3)\,|\,\xi(r_2),
    \xi(r_1)=0)$ as symbols.  $\sigma_{\infty}$ is the standard deviation of the velocity values $u_i$, or respectively $q_i$.
    The conditional PDF shows two points clearly. First, the dependency
    on the condition is seen in part (a) by the change with $x$-axis, this
    shows that $p(\xi(r)\,|\,\xi(r')) \neq p(\xi(r))$ or a  two-point closure is not supported.
   Considering features of the increments for only one selected scale, like done by  investigating
   structure functions, see Equation \ref{eq:multifrac}, are incomplete. 
   Second, the conditional PDF do not depend on a second condition on an increment on an even larger scale $r''$, 
   $p(\xi(r)\,|\,\xi(r'), \xi(r''))= p(\xi(r)\,|\,\xi(r'))$.
   This is taken as an indication that the $r$-evolution of
    the increments is memoryless, i.e., the knowledge of the value of
    the increment $\xi(r)$ suffices to determine the next step by which the
    system evolves to $\xi(r')$, with $r''<r' < r$. }
  \label{Fig_Markov}
\end{figure}

Having data from concrete complex systems, the validity of the
simplifications can be tested. As we focus in this contribution on the
reduction to three-point statistics, see Equation (\ref{eq:markov}), 
the validity of the simplification can be seen by investigating
$p(\xi_i|\xi_{i-1},\ldots,\xi_0)$. This is easily done by determining
different increments for the same reference value $q_N$. As an example the results from
one turbulent data set is shown in Fig. \ref{Fig_Markov}. 
Here the quantity of the system is the local velocity in the direction of the
mean flow, thus $q_i = u_i$. It can clearly be seen that
$p(\xi_3|\xi_2,\xi_1)$ depends on $\xi_2$, as the contour lines are not
parallel to the $\xi_2$-axis. Thus the
simplification of Equation (\ref{eq:two_point}), namely the reduction to two-point statistics, does not hold. It
can, however, be seen that the double conditioned PDF
$p(\xi_3|\xi_2,\xi_1)$  shown as red contours, is similar to the
single conditioned PDF $p(\xi_3|\xi_2)$. This result is a good
indication that Equation (\ref{eq:markov}), the three-point closure, holds. If many data are
available, conditions on further larger-scale increments can be
investigated. The quality of how well this condition is fulfilled can be
tested by statistical tests, see for example \cite{Friedrich1997,Renner2001}
and has been found for many data sets of turbulence \cite{Friedrich1998,Tutkun2004,Lueck2006}
and other data like financial data (e.g. \cite{Nawroth2010}) and surface heights, see \cite{Friedrich2011}. 
In \cite{Stresing2010} it is shown that this simplification holds also for $p(\xi_3|\xi_2,\xi_1,q_N) = p(\xi_3|\xi_2,q_N) $
and that for turbulence data $p(\xi_3|\xi_2,\xi_1,q_N)$ depends on the reference values $q_N$, too.
(Another way of showing Equation (\ref{eq:markov}) for experimental data is given by \cite{Marcq2001}).
Note that sometimes we observe that  both $p(\xi_3|\xi_2,\xi_1,q_N) = p(\xi_3|\xi_2,q_N) $
and  $p(\xi_3|\xi_2,\xi_1) = p(\xi_3|\xi_2) $ hold, but is it not a trivial point how these Markov conditions are related.
As mentioned for turbulence the Markov properties are found for both the multi-point $p(\xi_3|\xi_2,\xi_1,q_N)$ and the multi-scale
$p(\xi_3|\xi_2,\xi_1)$ statistics. For surface waves we found that Markov properties are only valid for the multi-point statistics \cite{Hadjihosseini2016}.
If the joint probabilities $W(\xi_3,\xi_2,\xi_1,q_N) $ can be written as $W(\xi_3,\xi_2,\xi_1,q_N)= W(\xi_3,\xi_2,\xi_1)W(q_N) $, one Markov property always follows from the other.

\begin{textbox}[h]
\section{WAVELETS, INCREMENTS, AND CORRELATIONS}

The description of the complex structure by multi-increment
statistics can be set in analogy to a wavelet analysis,
cf.~\cite{muzy1993multifractal,amblard1999cascade,farge2006wavelets,lovejoy2013weather}. 
Wavelets $\psi_{a,b}(x)$ are characterised by a scale $a$ (width) and a location $b$. Using the 
difference of two Dirac functions $\delta (x)$, the scale  $a=r_i$ and the loaction $b= x_N$ it is possible  to define the following wavelet
  $$ \psi_{r_i,x_N}(x) = \delta(x_N-x) - \delta (x_N-r_i-x), $$
  also sometimes called poor man's wavelet.
  Increments are nothing else than the coefficients of these wavelets
  $$ \xi_i= \xi(x_N,r_i)= \int_{-\infty}^{\infty}\psi_{r_i,x_N}(x)q(x) dx . $$
  
As discussed in \cite{amblard1999cascade}, we analyse the evolution of these coefficients of the wavelets with scales as stochastic processes.
In principle the discussion in this paper can 
also be performed with general wavelets $\psi_{r,x}(\cdot)$ and their coefficients $\tilde{\xi}(x,r)$. A main
difference is that the increment statistics can be related directly 
  to $N$-point statistics. For instance the correlation functions are given by the second oder moment 
  $\langle\xi^2(x,r)\rangle = 2\langle q^2\rangle -2\langle
  q(x-r)q(x)\rangle$. 
  Consequently, higher-order and mixed-order correlations are directly
  related to higher-order moments of the increment PDF, such as
  $\langle\xi^n(x,r)\rangle$.

\end{textbox}



\section{FOKKER-PLANCK EQUATION IN SCALE}
\label{Sec:FokkerPlanck}

So far the characterisation of complex disordered structures has been discussed by multi-point and multi-scale statistics, as well as the possible simplifications of three- and two-point closures. Next,
the hierarchical  ordering of the increments shown in Fig.~\ref{const_inc} together with the simplification of a three-point closure, which is achieved by Equation (\ref{eq:markov}), is put in the context of cascade processes, for which we work out a description by stochastic differential equations. Thus the aim is to grasp the whole complexity by some stochastic equations.

The basic first idea is to look at the increment for a chosen location $x_N$, i.e. $\xi (x_N,r_i)$ as a quantity that changes with $r$ and denote it by the increment trajectory 
$\xi (\cdot) \equiv \xi (x_N,r)$, which describes the above mentioned zoom-in process. Equation (\ref{eq:markov}) is now
nothing else than that $\xi (x_N,r_i)$ depends only on the increment of the next
larger scale $\xi (x_N,r_{i-1})$. At the same time $\xi (x_N,r_i)$ is independent of further increments on larger
scales. This means that the evolution of $\xi (\cdot)$ has no memory, note, this is the definition of a Markov process in scale $r$.
Thus this evolution has Markov properties and can be considered as a stochastic process
evolving in $r$, or more precisely, based on our definition, evolving
with decreasing $r$. (For readers who are not familiar with stochastic processes, we refer to the sidebar \ref{Sidebar:Markov} and the   continuative literature  \cite{Risken1984,hanggi1982,Gardiner1998}.)


\begin{textbox}[h]
\section{MARKOV PROCESS: FOKKER-PLANCK, KOLMOGOROV AND LANGEVIN EQUATION}
\label{Sidebar:Markov}
For a process of the quantity $\xi_i$ running from large to small scales, i.e. $i=N-1$ to $0$ with $r_i > r_{i+1}$,
the Markov processes are defined by the condition that 
 \begin{equation*}
 p(\xi_i|\xi_{i-1};\xi_{i-2};\cdots;\xi_0)=p(\xi_i|\xi_{i-1}).
 \end{equation*}
This means that conditional PDF depends only on the value $\xi_{i-1}$ at the closest scale.
For such Markov processes we can write
 \begin{eqnarray*}
W(\xi_i;\cdots;\xi_0) =  p(\xi_i|\xi_{i-1}) ~ W(\xi_{i-1};\cdots;\xi_0).
 \end{eqnarray*}
(Note the discussion presented here is in the same way valid, if the reference value $q_N$ is taken into account, too.) Using the same argument for $W(\xi_{i-1};\cdots;\xi_0)$, we find the following relation for the $i+1-$point joint PDF of Markov processes, 
  \begin{eqnarray*}
 W(\xi_i;\cdots;\xi_0)=p(\xi_i|\xi_{i-1}) \cdots p(\xi_1|\xi_0) ~  W(\xi_0) . 
 \end{eqnarray*}
Therefore marginal PDF $W(\xi_1)$ and conditional PDF $p(\xi_{k}|\xi_{k-1})$ are sufficient to describe Markov processes. The probability distributions (marginal and conditional) of Markov processes satisfy a partial differential equation of order one in the scale and order infinity in the state variable $\xi$. The governing equation is known as Kramers-Moyal (KM) equation, see Equation \ref{eq:km}.

In this respect the Pawula theorem states that there are only three possible cases in the KM expansion: \\
 i) the  Kramers-Moyal expansion stops at n = 1 means that the processes are deterministic,\\
  ii) the KM expansion stops at n = 2, the resulting equation is the Fokker-Planck or Kolmogorov equation and describes diffusion processes and finally \\
  iii) the Kramers-Moyal expansion stops at $n = \infty$. Any truncation of expansion at finite order $n > 2$ would produce non-positive probability density $W(\xi)$ \cite{Risken1984}. 

For the case (ii) the KM expansion reduces to the Fokker-Planck equation, which means that the first and second KM coefficients $D^{(1)}(\xi,r)$ (drift coefficient) and $D^{(2)}(\xi,r)$ (diffusion coefficient) are non-vanishing, see Equations \ref{eq:FPqN},\ref{eq:FP_sim}.  Now one can ask which dynamical equation governs the stochastic variable $\xi$ itself, where its marginal and conditional PDFs satisfy the Fokker-Planck equation. The corresponding stochastic equation is known as Langevin equation. Using the It\^o interpretation it has the following form \cite{Friedrich2011}
\begin{equation*}
-r \frac{d \xi}{dr}= D^{(1)}(\xi,r) + \sqrt{D^{(2)}(\xi,r)}\eta(r)  ,
\end{equation*}
where noise $\eta(r)$ is a zero mean white Gaussian with intensity $2$, which means that $\langle \eta(r) \eta(r') \rangle = 2 \delta(r-r')$.  

\end{textbox}


\begin{textbox}[h]
\section{COMMENTS ON MARKOV PROCESSES}

\subsection{direction of the process}
The evolution of the scales from large to small values is considered, for this ourpose we used in Equation \ref{eq:km} a negative prefactor of $-r$. Having $-r /\partial r = - 1/\partial ln(r) $  shows that we have implicitly used a log scaling of the $r$-evolution, which is of advantage for complex structures with self-similar properties, see  Chapter \ref{Sec:fractal}.
\subsection{inverse direction of the process}
If it is shown that the data fulfill the Markov conditions Equation \ref{eq:markov} from large to small scales, 
the Markov condition is fulfilled also in the other direction, from small to large scales \cite{RennerDiss}  
\begin{equation}
p(\xi_i|\xi_{i+1},...,\xi_{N-1},q_N) = p(\xi_i|\xi_{i+1},q_N)  
\label{eq:markov_inv}
\end{equation}
as 
\begin{eqnarray}
p(\xi_i|\xi_{i+1},...,\xi_{N-1},q_N) &=& \frac{W(\xi_i,\xi_{i+1},...,\xi_{N-1},q_N)}{W(\xi_{i+1},...,\xi_{N-1},q_N)} 
= \frac{W(\xi_{N-1},\xi_{N-2},...,\xi_{i},q_N)}{W(\xi_{N-1},\xi_{N-2},...,\xi_{i+1},q_N)} \\
&=& \frac{p(\xi_{N-1}|\xi_{N-2},q_N)  ... p(\xi_{i+2}|\xi_{i+1},q_N) \; p(\xi_{i+1}|\xi_{i},q_N) W(\xi_i,q_N)}{p(\xi_{N-1}|\xi_{N-2},q_N)  ... p(\xi_{i+2}|\xi_{i+1},q_N) W(\xi_{i+1},q_N)} \\
&=& \frac{W(\xi_{i+1},\xi_i,q_N)}{W(\xi_{i+1},q_N)} = p(\xi_i|\xi_{i+1},q_N)  .
\end{eqnarray}

\end{textbox}

For experimental or empirical data like from turbulence or from 
financial markets \cite{Renner2001finance} the absence of memory gets lost for the smallest scales.
Such a behaviour has already been proposed as a natural one by Einstein in his pioneering work in
Brownian motion \cite{Einstein1905}. This defines a lower bound scale, 
which we call Einstein-Markov length $r_{EM}$ and which can be determined by 
the validity of Equation (\ref{eq:markov}) as $x_{N-2}$ converges against
$x_{N-1}$ \cite{Lueck2006}. Thus our consideration is valid for $r>r_{EM}$ 
and may be treated as a small scale cut-off, see e.g. \cite{Dubrulle2000}. Note that $r_{EM}$
is more than a lower bound, but is also the finite step size that coarse-gains the whole $r$- evolution 
from largest to smallest scale. In this way the Markov process may be taken as a 
stochastic process modelling in a continuous manner the coarse-grained process.

The evolution of the cPDFs $p(\xi|\xi_i,q_N)$ with $r<r_i$ describes the transition probability 
of $\xi_i(r_i) \rightarrow \xi(r)$ for the given reference $q_N$. An equation for the evolution of this
transition probability with extending the difference between $r$ and $r_i$ is given by the Kramers-Moyal expansion
(more precisely the Kramers-Moyal forward expansion) \cite{Risken1984}

\begin{equation}
  -r\frac{\partial}{\partial r}p(\xi|\xi_i,q_N)=
  \sum_{n=1}^{\infty}\left(-\frac{\partial}{\partial \xi}\right)^n
  \left[ D^{(n)}(\xi,r,q_N)p(\xi|\xi_i,q_N)\right] .
\label{eq:km}
\end{equation}
$D^{(n)}$ are called Kramers-Moyal coefficients and can be found from time series, see below. If the fourth order Kramers-Moyal coefficient $D^{(4)}$ vanishes, the Kramers-Moyal expansion reduces after Pawula's Theorem, cf.~\cite{Risken1984} to a Fokker-Planck equation, which is also known as Kolmogorov equation \cite{Kolmogorov1931}.
For the Fokker-Planck equation the expansion of Equation (\ref{eq:km}) truncates after the second term 
\begin{equation}
-r\frac{\partial}{\partial r}p(\xi|\xi_i,q_N)= -\frac{\partial}{\partial \xi} \left[ D^{(1)}(\xi,r,q_N)p(\xi|\xi_i,q_N)\right] + \frac{\partial^2}{\partial \xi^2} \left[ D^{(2)}(\xi,r,q_N)p(\xi|\xi_i,q_N)\right]. 
\label{eq:FPqN}
\end{equation}
Note that if $p(\xi|\xi_i,q_N)$ is independent on $q_N$ and if one multiplies the equation with $W(\xi)$ and integrates over $\xi_i$, Equation~(\ref{eq:FPqN}) reduces to
\begin{equation}
-r\frac{\partial}{\partial r}W(\xi)= -\frac{\partial}{\partial \xi} \left[ D^{(1)}(\xi,r)W(\xi)\right] + \frac{\partial^2}{\partial \xi^2} \left[ D^{(2)}(\xi,r)W(\xi)\right],
\label{eq:FP_sim}
\end{equation}
which leads to the pure description of the increment statistics of $W(\xi)$. This is clearly 
less information than $p(\xi|\xi_i)$, which we know from the discussion of the multi-point statistics in the previous section.
Note that $D^{(n)}(\xi,r,q_N)$ and $D^{(n)}(\xi,r)$ are related by  $\int D^{(n)}(\xi,r,q_N)p(\xi|\xi_i,q_N)W(q_N) d q_N ={D^{(n)}}(\xi,r)p(\xi|\xi_i) $.

Looking at the evolution of the increments $\xi_r$ with scale, the Markov property of 
Equation (\ref{eq:markov}) means that only delta correlated noise acts on the trajectory.
The reduction of the Kramers-Moyal expansion of Equation (\ref{eq:km})
goes along with the requirement that the involved noise in the
stochastic process is not only delta correlated but has also Gaussian
distribution. This is also called Langevin noise, for
which a corresponding differential equation for a single event or path
$\xi(\cdot)$ is given as
\begin{equation}
-r\frac{\partial}{\partial r} \xi = D^{(1)}(\xi,r)  + \sqrt{ D^{(2)}(\xi,r) } \eta(r). 
\label{eq:langevin}
\end{equation}
where $\eta(r)$ denotes zero-mean Gaussian white noise with a variance of $2$, i.e.
$\langle \eta(r) \eta(r') \rangle = 2 \delta(r-r')$. Here we use the It\^o interpretation. For Stratonovich and other descriptions see cf.~\cite{Risken1984,hanggi1982}. From the Langevin equation it is evident that $D^{(1)}$ describes the deterministic part of this equation and is called drift coefficient. The function $D^{(2)}(\xi,r)$, which is called diffusion coefficient, determines the amplitude of the noise. The case that $D^{(2)}$
changes with $\xi$ is called multiplicative noise.


An essential point for the stochastic description of the scale
dependent increments is the knowledge of the Kramers-Moyal
coefficients $D^{(n)}$, which can be determined directly from the data
as conditional moments (cf.~\cite{Risken1984,Friedrich2011}).

First let us define the n-th order moments for two increments in scales that are separated by $\delta$
\begin{eqnarray}
M^{(n)}(\delta,\xi,r, q_N)  &=&   \left\langle \left[\xi^{\prime}(r-\delta,q_N)-\xi(r,q_N)\right]^n \right\rangle|_{\xi(r,q_N)={\xi} } \\
  &=&  \int_{-\infty}^{\infty} \left(\left[ \xi^{\prime}(r-\delta,q_N)-\xi_j(r,q_N)\right]^n\right) p(\xi^{\prime}|\xi,q_N) d \xi^{\prime} .
\label{eq:km_mom}
\end{eqnarray}
The  values of $M^{(n)}$ depend on the value of $\delta$ for some chosen or fixed values of $\xi,r, q_N$. The Kramers-Moyal (KM) coefficients $D^{(n)}$ are given by
\begin{eqnarray}
  D^{(n)}(\xi,r, q_N)
    &=& \frac{r}{n!} ~ \lim_{\delta \to 0} \frac{1}{\delta} M^{(n)}(\delta,\xi,r, q_N).
\label{eq:km_coef}
\end{eqnarray}
The definition presented here for KM coefficients differ by the factor of $r$ with common definition, which is due to our description of a stochastic process in scale, see Equation (\ref{eq:langevin}).
It should be mentioned that this definition of the Kramers-Moyal coefficient can already be found in an early work of Kolmogorov in 1931 \cite{Kolmogorov1931}.

We see that the cPDFs, $p(\xi^{\prime}|\xi,q_N)$ play again an important role,
as for the $\lim_{\delta \to 0}$ the differential equation (\ref{eq:km})
can be estimated from their knowledge. Note that this
$\lim_{\delta \to 0}$ can be considered as a fusion process of two
increments $\xi_j^{\prime} \rightarrow \xi_j$, which is of interest for a field
theoretical approach to such complex systems \cite{Friedrich2011,Reza2000,FriedrichJDiss}.

\begin{figure} 
  \centering
  \includegraphics[width=0.8\linewidth]{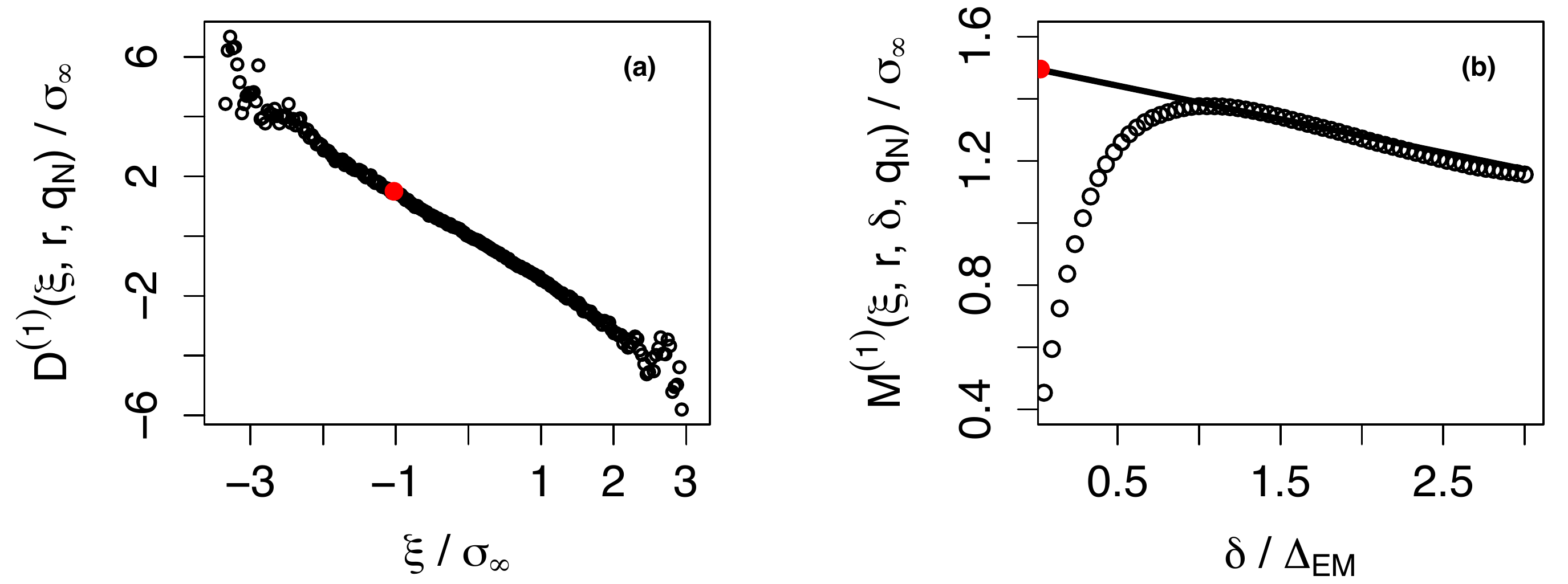}
  \caption{The scheme how Kramers-Moyal coefficients can be
    determined using Equation \ref{eq:km_coef} is illustrated for
    the  first Kramers-Moyal coefficient, or drift term. (a) shows the
    drift term for a turbulent flow \cite{Renner2001}. The red dot of
    $D^{(1)}$ marks a value for which the conditional moment
    $M^{(1)}(\xi,r,\delta, q_N)$ is evaluated, see figure part (b).
    The function $D^{(1)}(\xi,r,q_N)$ is obtained point by point using
    a linear extrapolation of $M^{(1)}(\xi,r,\delta, q_N)$ for
    $\lim_{\delta \to  0}$, shown as a solid line (b).
    Note the linear shape of $D^{(1)}$ reflects the deviation of
    conditional PDF from the diagonal shown in Fig.~\ref{Fig_Markov}a,
    and means for the evolution of the increments, after
    equation~\ref{eq:langevin}, that their sizes decrease as the
    scales $r$ become smaller. }
  \label{Fig_KM}
\end{figure}

An important aspect of Equations \ref{eq:km_mom},\ref{eq:km_coef}  is that they can also be read as a concept to estimate the Kramers-Moyal coefficients directly from given data, as shown for some turbulence data in Fig. \ref{Fig_KM}. That this is a very efficient method to analyse time series of noisy dynamical systems has been shown in \cite{Friedrich2011}.
Experiences have shown that technically the limit $\lim_{\delta \to  0}$ is best performed by investigating the moments
$M^{(n)}(\xi,r,\delta, q_N) $
with the help of the small step approximation in $\delta$, for which
$M^{(n)}(\xi,r,\delta, q_N) = D^{(n)}(\xi,r, q_N) \delta +
\mathcal{O}(\delta^2)$, see Fig. \ref{Fig_KM}b. If the given data do not allow this small step
approximation, due to insufficient sampling rates or due to a too large Einstein-Markov
length, respective corrections can be calculated  \cite{Gottschall2008, Honisch2011}. In  Fig. \ref{Fig_KM}b the sampling
rate was sufficiently high. The deviation from a linear law for small values $\delta$ is due to the Einstein-Markov length. 
Other corrections arise, if additionally measurement or observation noise or
another non-ideal noise contribution is given cf. \cite{siefert2003quantitative,bottcher2006reconstruction,lehle2011analysis,lehle2018analyzing}.

\begin{figure}
  %
  %
  \includegraphics[width=1.0\linewidth]{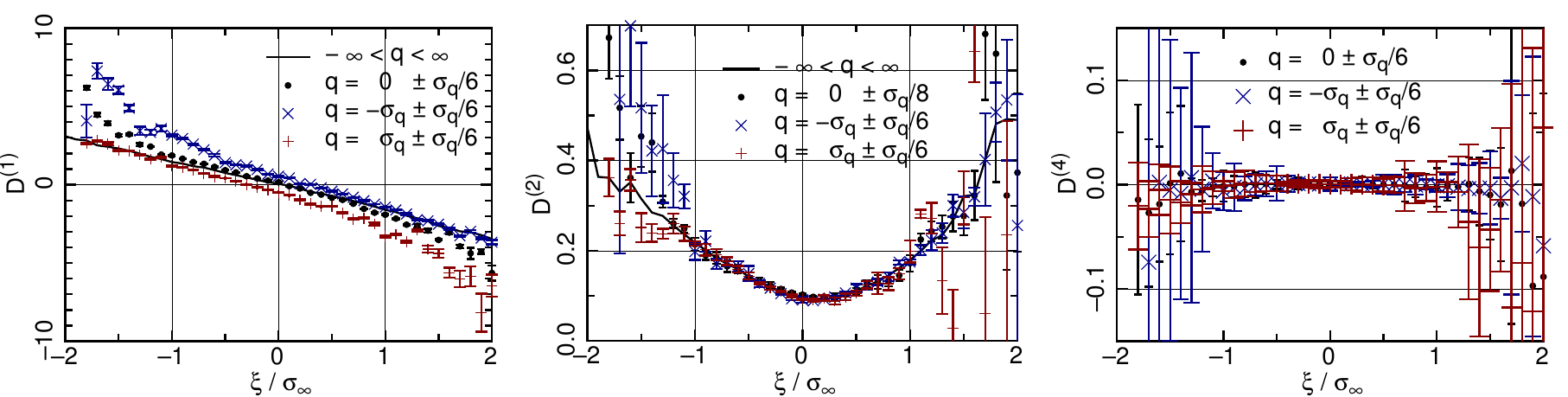}
  \caption{First, second and fourth order Kramers-Moyal
    coefficients $D^{(n)}(\xi,r,q_N)$ calculated from data of a
    turbulent flow, using the procedure illustrated in Fig.
    \ref{Fig_KM}. Three different reference values $q_N$, labeled
    as
    $q=-\sigma_q \pm \sigma_q/6,\; q= 0 \pm \sigma_q/6,\;
    q=\sigma_q \pm \sigma_q/6$, were chosen. To get sufficient
    data for the statistics, always a small but finite interval
    around the $q_N$ values was chosen. Here $\sigma_q$ is the
    standard deviation of $W(\xi)$. Note that only
    $D^{(1)}(\xi,r,q_N)$ shows clear dependence on $q_N$. For
    further details see~\cite{Stresing2010}. The negative values
    for the fourth order Kramers-Moyal coefficient are due to the
    extrapolation procedure as indicated in Figure~\ref{Fig_KM}.
    Once these Kramers-Moyal coefficients are successfully estimated
    as in this figure, the corresponding Fokker-Planck equation is
    completely obtained. }
  \label{Fig_KM_qN}
\end{figure}

In Fig. \ref{Fig_KM_qN} the Kramers-Moyal coefficients are shown for a turbulent data set. The linear behaviour of the drift term $D^{(1)}$ and the quadratic behaviour of the diffusion term $D^{(2)}$ becomes clear. $D^{(2)}$ has an additional additive offset. The fourth order Kramers-Moyal coefficient can be taken as zero within the experimental precision. This indicates that for turbulence a Fokker-Planck equation can be used to describe the cascade  process and thus the whole multi-point statistics. For the case that $D^{(4)}$ does not become zero, in principle infinitely many Kramers-Moyal coefficients  have to be determined. A criterium to quantify the importance of higher order Kramers-Moyal coefficients  has been worked out in \cite{Renner2001}, and in \cite{Tutkun2004} it has been shown for turbulence that in contrast to the velocity fields the passive scalars require such higher order Kramers-Moyal coefficients. Another aspect of the results for the Kramers-Moyal coefficients shown in Fig. \ref{Fig_KM_qN} is that the dependency on the reference point $q_N$ is only clearly present for $D^{(1)}$. For positive $q_N$  the fixed point $D^{(1)}(\xi) = 0$ is shifted to negative values, whereas for negative $q_N$ the fixed point is shifted to positive values. This result simply means that for positive $q_N$ the increments have the tendency to become more negative, what is in accordance with the boundedness of turbulent velocity data from stationary experiments. 

Knowing the coefficients $D^{(1)}$ and $D^{(2)}$ the cPDF $p( \xi_r | \xi_{r +\delta})$ (as well as for $p( \xi_r | \xi_{r +\delta},q_N)$ ) can be
calculated by the short time propagator 
\cite{Risken1984,Renner2001}
\begin{eqnarray}
p_{STP}( \xi_r | \xi_{r +\delta})&\approx& \frac{1}{\sqrt{4\pi  \;D^{(2)}( \xi_{r + \delta} )\;  \delta  }}\nonumber\\
&\times& \exp\left(-\frac{\lbrack \xi_{r}-\xi_{r+\delta} -D^{(1)} ( \xi_{r +\delta} ) \delta \rbrack^2}{4 D^{(2)}(\xi_{r +\delta})\delta }\right) \quad.
\label{eq:short_time_prop}
\end{eqnarray}
A sufficiently small step $\delta$ has to be used to stay in this limiting approximation. Based on this short time
propagator any cPDF $p( \xi_r | \xi_{r'})$ can be determined. 
Therefore the quality of the estimation of the coefficients $D^{(1)}$ and $D^{(2)}$ can be verified by comparing the
cPDFs obtained from the data with those obtained by solving
the Fokker-Planck equation with the estimated coefficients \cite{Nawroth2007}.


\begin{textbox}[h]
\section{HIGHER-ORDER KRAMERS-MOYAL COEFFICIENTS  }

  According to the Pawula theorem vanishing higher order Kramers-Moyal (KM) coefficients, specially fourth-order $D^{(4)}(\xi,r)$, guarantee
that the process is statistically contineous and the Kramers-Moyal expansion Eq.(16)
 can be truncated after the
second (diffusive) term cf.~\cite{Risken1984}. For vanishing KM coefficients one can construct the
Langevin equation with the computed drift function and diffusion coefficients form time series.
 Non-vanishing higher-order ($n>2$) KM coefficients have been observed
in various systems cf.~\cite{Friedrich2011}, which indicates that
the corresponding measured time series do not belong to the class of continuous
diffusion processes and jump events should play a significant role
in the underlying stochastic process \cite{Anvari2016}.
  
\end{textbox}



\section{SELF-SIMILARITY AND FRACTALS}
\label{Sec:fractal}

With the previous Section~\ref{Sec:FokkerPlanck} the derivation of a
Fokker-Planck equation as a model of scale-dependent complexity has
been completed. We now want to put this approach in the context of
other analysis frameworks for complex systems, namely self-similarity
and fractals.
For complex structures the question is often posed, whether they possess self-similar structures, also called fractals. Particularly for the two examples of turbulence and sea waves, discussed here, the concept of self-similarity plays an important role, cf. \cite{Frisch2001,Nazarenko2016}.  We start the discussion of self-similarity in a quite general way with the principles of scaling  symmetries, from which we derive properties of the so-called structure functions $\langle \xi^n \rangle$. The structure functions are very often used for the characterisation of both turbulence and surface roughness like sea waves. We show how the concept of the stochastic cascade equations of the previous Section can describe these structure functions and their self-similiar structure.

 Commonly the self-similarity is
investigated by a local measure, which characterises the structure on the scale $r$ at the location $x$. 
We denote the local measure again as $\xi(x,r)$. 
Self-similarity means that in a certain range of $r$ the quantities
\begin{equation}\label{scalinv}
\xi(x,r) \qquad , \qquad \lambda^\alpha  \xi(\lambda r, \lambda^\beta x)
\end{equation}
should have the same statistics. More precisely, 
the probability distribution of the quantity $\xi$ takes the form
\begin{equation}\label{similar}
W(\xi,r)=\frac{1}{r^\alpha} F(\frac{\xi}{r^\alpha})
\end{equation}
with a universal function $F(Q)$. The universality of $F$ leads to the
scaling behaviour 
\begin{equation}\label{eq:fractal}
\langle\xi^k(r)\rangle=\int  \xi^k \frac{1}{r^\alpha}
F(\frac{\xi}{r^\alpha})  \quad d\xi  =Q_k r^{k \alpha}  .
\end{equation}
Such a type of behaviour has been termed {\em fractal} scaling behaviour.

The concept of fractals is widespread and many examples are known, like 
turbulence or surface roughness, just to mention two. 
The strict self-similarity expressed by Equation (\ref{scalinv}) is often just
an idealised approximation. 
In fact the so-called {\em multifractal behaviour} is often more appropriate. Here 
 the k-th order moments scale according to
\begin{equation} \label{eq:multifrac}
\langle\xi^k(r)\rangle =Q_k r^{\zeta(k)} \qquad ,
\end{equation}
where the scaling indices $\zeta(k)$ are now not any more linear but a
nonlinear function of the order $k$.

Such a multifractal behaviour can formally 
be obtained by the assumption that the probability
distribution $W(\xi,r)$ has the following form
\begin{equation}
W(\xi,r)= \int  \tilde{W}(\alpha,r) \frac{1}{r^{\alpha}}F(\frac{\xi}{r^\alpha}) \; d\alpha  \quad .
\end{equation}
This formula is based on the idea that the complex system is composed of 
subsets of different scaling indices $\alpha$,
where $\tilde{W}(\alpha,r)$ gives a measure of the scaling indices $\alpha$ at a scale $r$ (see e.g. \cite{Castaing1990,Castaing1996} ).
A shortcoming of the fractal and multifractal approach to complexity in scale
is the fact that it only addresses the statistics of the measure $\xi(x,r)$
at a single scale $r$. As we have derived above, 
one has to expect for a general N-point characterization dependencies
of the measures $\xi(x,r)$ and $\xi(x,r')$ from different scales, as well as dependencies in the value
of a reference point $q(x)$. 

The connection between the fractal and multifractal characterisation and the stochastic cascade description
can be derived from the Kramers-Moyal expansion of Equation (\ref{eq:km}). The validity of the Markov property or, 
respectively, the three-point closure is assumed and the Equation (\ref{eq:km}) has been integrated over all
values of $q_n$, so that the dependency on the reference value is not any more taken into account
\begin{equation}
  -r\frac{\partial}{\partial r}p(\xi|\xi_i)=
  \sum_{n=1}^{\infty}\left(-\frac{\partial}{\partial \xi}\right)^n
  \left[ D^{(n)}(\xi,r,q_N)p(\xi|\xi_i)\right] .
\label{km2}
\end{equation}
The multiplying of this equation with $\xi^k$ and 
the partial integration over $\xi$, e.g. \cite{Renner2001}
 leads to 
\begin{equation}\label{DGStrFkt}
   -r \frac {d} {d r} \langle(\xi_r)^k\rangle = 
   \sum_{n=1}^{k-1} \frac {k !} {(k-n)!} \langle D^{(n)} \xi_r^{k-n}\rangle .
\end{equation}
 If the Kramers-Moyal coefficients have the form $D^{(n)} = d_{n} \xi^n$
 (where $d_{q}$ are constants) \cite{Friedrich1997c}, scaling behaviour of (\ref{eq:multifrac}) is
 guaranteed with 
\begin{equation} \label{zeta_n}
\zeta_{k}= -\sum_{n=1}^{k-1} \frac {k !} {(k-n)!}
 d_{n} .
 \end{equation}
 Based on this formula it can be worked out which combinations of $d_n$ will result in the
$q$-dependent function of $\zeta_{q}$, which characterizes different multifractal models.

For turbulence and increments $\langle(\xi_r)^k\rangle$ is a common quantity to characterise different 
flow situations. $\langle(\xi_r)^k\rangle$  is called the $k$th order structure function.
If the Kramers-Moyal expansion truncates to a Fokker-Planck equation the structure functions can be found from the following equation
\begin{eqnarray} 
\label{eq:momentsFP}
	-r \frac{\partial}{\partial r} \langle \xi_r^k \rangle =  k \langle \xi_r ^{(k -1)} D^{(1)}(\xi_r) \rangle  + r k (k-1) \langle \xi_r^{(k -2)}  D^{(2)}(\xi_r) \rangle
\end{eqnarray}
From Fig. \ref{Fig_KM} it is obvious that  the drift coefficient has a linear  behaviour, $D^{(1)}( \xi_r) = d_{11} \; \xi_r$, and  the diffusion coefficient has a quadratic behaviour,  $D^{(2)}( \xi_r) = d_{20} + d_{22}\; \xi_r^2$. 
The $ d_{ij}$ may be r-dependent. The scaling index becomes now
\begin{eqnarray} 
\label{eq:momentsFP2}
\zeta_{k} &=& \frac{r}{\langle\xi_r^k\rangle} \frac{\partial \langle\xi_r^k\rangle}{\partial r} 
     = \frac{\partial \ln(\langle\xi_r^k\rangle)}{\partial{\ln(r)}}  \\
    &=& 
- k \biggl( d_{11}(r)   + (k-1) \Bigl(  d_{22}(r)+  \frac{\langle\xi_r^{k-2}\rangle}{\langle\xi_r^k\rangle}d_{20}(r) \Bigr)\biggr) . 
\end{eqnarray}
Due to the additive term in $D^{(2)}(\xi_r)$ a mixing of different structure functions with different orders takes place. For the
case of $d_{20}=0$, which is not supported by experimental data (see Fig. \ref{Fig_Markov}), and for constant values of 
$d_{11}$  and $d_{22}$, the so called Kolmogorov 1962 (K62) or lognormal model \cite{Kolmogorov1962} is obtained with the intermittency parameter $\mu$
\begin{eqnarray} 
\label{eq:K62}
\zeta_{k}  &=& -  k  d_{11}   + k(k-1)   d_{22}  \\
&=&  \frac{k}{3} - \mu \frac{k(k-3)}{18} .
\end{eqnarray}
Thus the K62 scaling corresponds to $d_{11} = -\frac{3+\mu}{9}$ and $d_{2} = \frac{\mu}{18}$.
The corresponding relation between stochastic processes and the other
well known multifractal scaling models of turbulence has been worked out  by \cite{NickelsenDiss,FriedrichJDiss}.



\section{SURROGATE DATA AND FORECASTING}
\label{Sec:sorrogate}

The Fokker-Planck equation (FPE) derived in
Section~\ref{Sec:FokkerPlanck}, despite its compactness, achieves a
comprehensive and powerful characterisation and description of a wide
range of complex systems. However, the generation of surrogate data
which fully obey a special FPE is not trivial. This section will
develop an approach to this task.

Based on the relation of the general $N$-point statistics of a complex structure  and a stochastic description by a Fokker-Planck equation
we obtain the possibility to generate new data sets numerically or to forecast special events. Therefore we consider the case that
 preceding values $q_0,\cdots,q_{N-1}$  fix the probability of a new value
 $q_N$. Taking this row of values $q_i, i=0,\cdots,N-1$ as a sequence of events, the cPDF of Equation (\ref{eq:c_pdf}) can be seen as the prediction of the next event for the next time step  (see e.g. \cite{Hallerberg2007}). Such a predictor can be expressed by the stochastic cascade process, using Equations (\ref{eq:Winc},\ref{eq:Winc_mar})

\begin{eqnarray} \label{eq:predic_qN}
  p(q_N| q_{N-1},\ldots,q_0)
  &=&  \frac{W(\xi_0,\xi_1,\ldots,\xi_{N-2},\xi_{N-1},q_N)}{W(\xi_0,\xi_1,\ldots,\xi_{N-2},q_{N-1})}  \\
 &=&     \frac{p(\xi_{N-1}|\xi_{N-2},q_N)\cdot  p(\xi_{N-2}|\xi_{N-3},q_N)   \cdot\ldots\cdot  p(\xi_{0}|q_N) \cdot W(q_N)}
    {p(\xi_{N-2}|\xi_{N-3},q_{N-1})\cdot  p(\xi_{N-3}|\xi_{N-4},q_{N-1})   \cdot\ldots\cdot    p(\xi_{0}|q_{N-1}) \cdot W(q_{N-1})} \\
        &=& \frac{\prod_{i=1}^{N-1} p(\xi_i|\xi_{i-1},q_N)}
            {\prod_{i=1}^{N-2} p(\xi_i|\xi_{i-1},q_{N-1})}
       \times
       \frac{p(\xi_{0}|q_N)}{p(\xi_{0}|q_{N-1})}
       \times
       \frac{W(q_N)}{W(q_{N-1})} \;.
\end{eqnarray}
Note that the increments $\xi_i$, defined in Equation (\ref{inc}),  have to be taken from the reference value $q_N$ in
the nominators and from reference value $q_{N-1}$ in the denominators, respectively.

Equation (\ref{eq:predic_qN}) enables to determine the probability of the new value $q_N$ based on the knowledge of the simple conditional PDFs $p(\xi_i|\xi_{j},q_N)$, which can either be calculated from the Fokker-Planck equation or which can be estimated directly from the data. As $p(\xi_i|\xi_{j},q_N)$ contains only knowledge of three values $q_i$, $q_{j}$ and $q_N$  of the data, this is again a three- point closure of multi-point statistics.

\begin{figure}
  \includegraphics[width=0.8\linewidth]{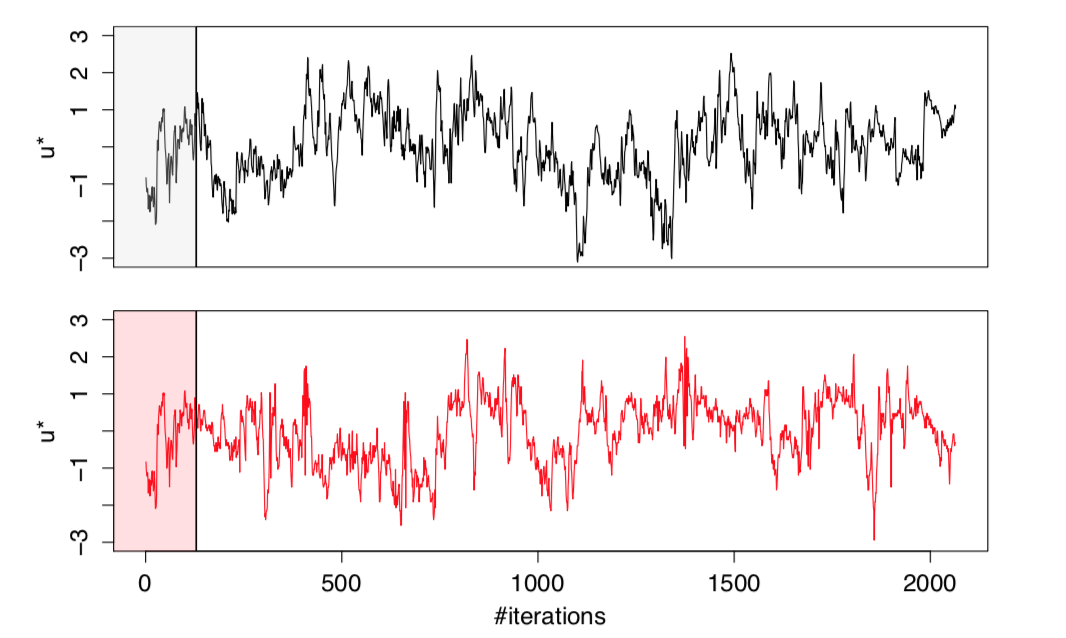}
  \caption{Time series of wind data as a real-world example of
    numerical data generation, cf.{}
    Sec.~\ref{Sec:sorrogate}. The upper panel shows the measured data,
    while the lower panel of the figure presents reconstructed data
    using Equation (\ref{eq:predic_qN}). The data, which are used as initial condition for the reconstruction, are marked 
    on the left sides of the panels 
    It is well visible that the reconstructed data are not identical with
    the measured data it can be shown that they but follows the same statistics. Not only mean
    value and standard deviation are correctly reproduced, but also
    multi-point statistics and higher-order correlations. }
  \label{Fig_recon_signal}
\end{figure}

The conditional probabilities $ p(q_N| q_{N-1},\ldots,q_0)$  contain all relevant statistical information of the previous data points for a correct choice of the new value $q_N$. Choosing now a random value from this distribution, the time series will be extended correctly by another point.
 Shifting the procedure by one step and repeating the same procedure may be used to generate new surrogate time series, which exhibits the correct joint probability density function for all considered scales.  For technical reasons one should avoid zeros in cPDFs of the denominator  of Eq. (\ref{eq:predic_qN}). The initial idea for reconstructing time-series following this procedure was  developed in a similar way for fluid turbulence data, see \cite{Nawroth2006}, and has been used for turbulent data  \cite{Stresing2010}, for financial data \cite{Nawroth2010}, and also for sea waves \cite{Hadjihosseini2016}. In Figure \ref{Fig_recon_signal} we show two time series of wind speed measurements. In the upper panel the originally measured time series is shown, in the lower panel a time series obtained by the just mentioned reconstruction method. The color-coded left part represents the initial conditions of the first N values $q_0,...,q_{N-1}$, used to start the reconstruction method.  As this is a stochastic model, involving a deterministic as well as a random part, the two time series diverge quite fast. But the stochastic content in the sense of multi-point statistics is the same, which can be verified by reanalyising these surrogate data \cite{Stresing2010,Nawroth2010,Hadjihosseini2016}. Another interesting point is that apparently typical structures of a wave pattern could be reproduced by the stochastic method \cite{Hadjihoseini2017}, thus it seems that the multi-point approach is capable to grasp the statistics as well as coherent structures. This will only work, if such structures are based on the special stochasticity and it will not work if special structures are added to a noisy background.

\begin{figure}[thb]
  \centering
  \includegraphics[width=\textwidth]{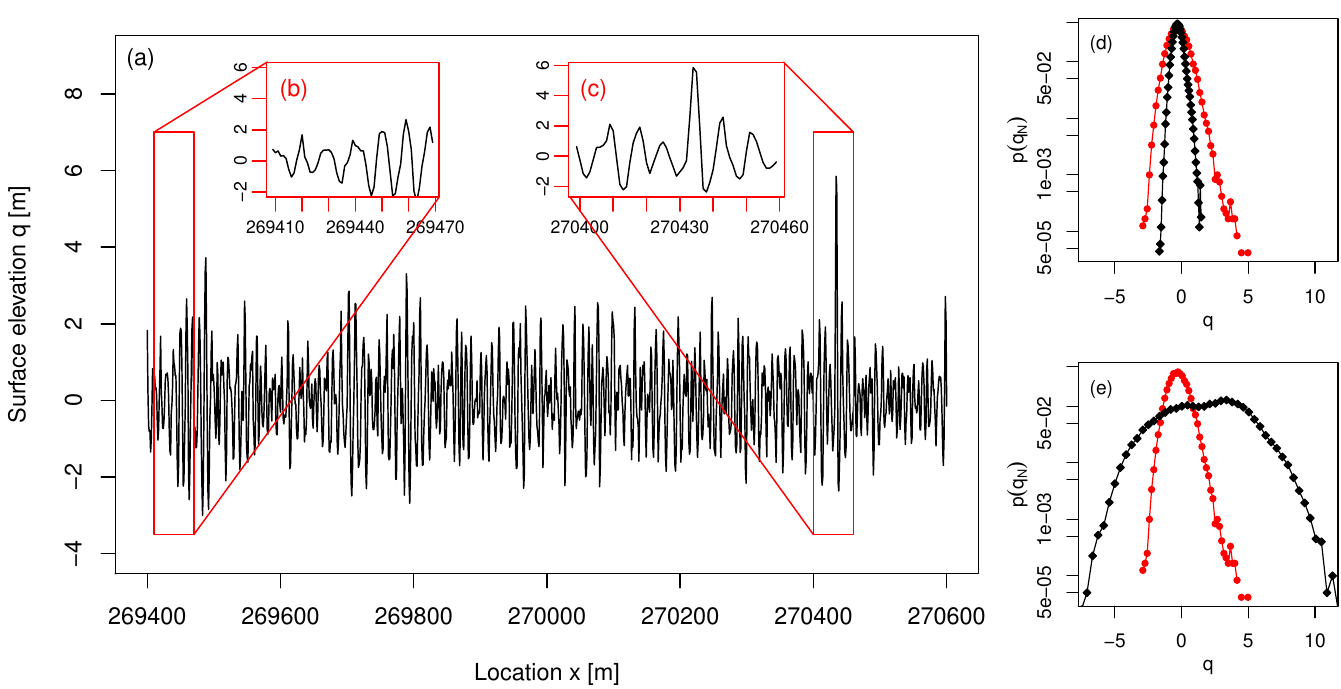}
  \caption{ Time series and PDF of ocean gravity waves after
    \cite{Hadjihosseini2016}, as a second example of real-world data
    reconstruction.
    (a) Reconstructed time series after Eq.~\ref{eq:predic_qN}. Two
    time windows are marked by (b) and (c) for which the
    corresponding multi-conditional PDFs (after equation
    \ref{eq:predic_qN}) are given in (d) and (e). To
    show the changing volatility, both the multi-conditional PDFs (black) as
    well as the unconditional PDFs (red), estimated from all data, are
    shown. Note the obvious changes of the likelihood of large
    wave amplitudes.
    It becomes clear how the multi-point statistics change along the
    time series, defining regions of smaller and larger wave
    amplitudes, respectively. As another aspect, the multi-conditional
    PDFs can serve for short time forecasting. For consistency the 
    heights of the waves are denoted by the variable $q$, and the time dependences has been
    transformed to a spatial dependency using a wave velocity of $1$m/s. }
  \label{recon_time_series}
\end{figure}

The method to reconstruct data sets with the conditional probabilities
$ p(q_N| q_{N-1},\ldots,q_0)$ can also be used for a short time
forecast, as it was shown for financial data \cite{Nawroth2010} and
sea waves \cite{Hadjihosseini2016}. In Fig. \ref{recon_time_series}(a)
typical time series of wave heights is shown. Note that the big wave
at the end of the time series corresponds to a measured rogue wave. In
the figure part (d) and (e) two selected conditional probabilities
$p(q_N|q_0,r_0, \dots, q_{N-1},r_{N-1})$
are shown to illustrate our
method.
In addition to the conditional probabilities
the single event probability
$p(q_N)= p(q)$ of all height values is shown (red curve). These
figures show clearly how the conditional probabilities change with
$q_0,r_0, \dots, q_{N-1},r_{N-1}$ the values of the $N$ wave heights
seen before. There are cases when smaller $q_N$ values are expected in
the next step, see Fig.~\ref{recon_time_series}(b), and there are
cases when large $q_N$ values become highly likely, see
Fig.~\ref{recon_time_series}(c). With this method a warning system for
approaching large wave heights can be set up. The high quality of such
a prediction was quantified according to the receiver operating
characteristic curve (ROC) \cite{Hadjihosseini2016}.

\section{NON-EQUILIBRIUM THERMODYNAMICS OF COMPLEX HIERARCHICAL STRUCTURES}
\label{Sec:Thermo}

Up to this point we have outlined a statistical approach to characterise completely the disordered structure in one direction by multi-point statistics. This approach was a phenomenological one. For the case of turbulence there are some works which show how such a Fokker-Planck equation can be related to the basic equations of fluid mechanics. In \cite{Laval2001} the connection to the Navier-Stokes equation is shown, in \cite{FriedrichJDiss} the Lundgren hierarchy was analysed based on a three-point closure or, respectively, Markov properties in scale. For the sea waves comparable results are not known to us.
In this last Section we put our statistical approach in the context of non-equilibirum thermodynamics (for earlier approaches to non-equilibrium approaches see for example \cite{brown1982information}). Based on the derived Fokker-Planck equations for the cascade process we can assign entropy values to each local structure of the complex systems. For these entropy values the validity of a fluctuation theorem, namely the integral fluctuation theorem, can be shown. This is a way for how the phenomenological stochastic approach can be linked to fundamental laws of physics, cf. \cite{Seifert2012}.

\begin{figure}
  \includegraphics[width=0.7\linewidth]{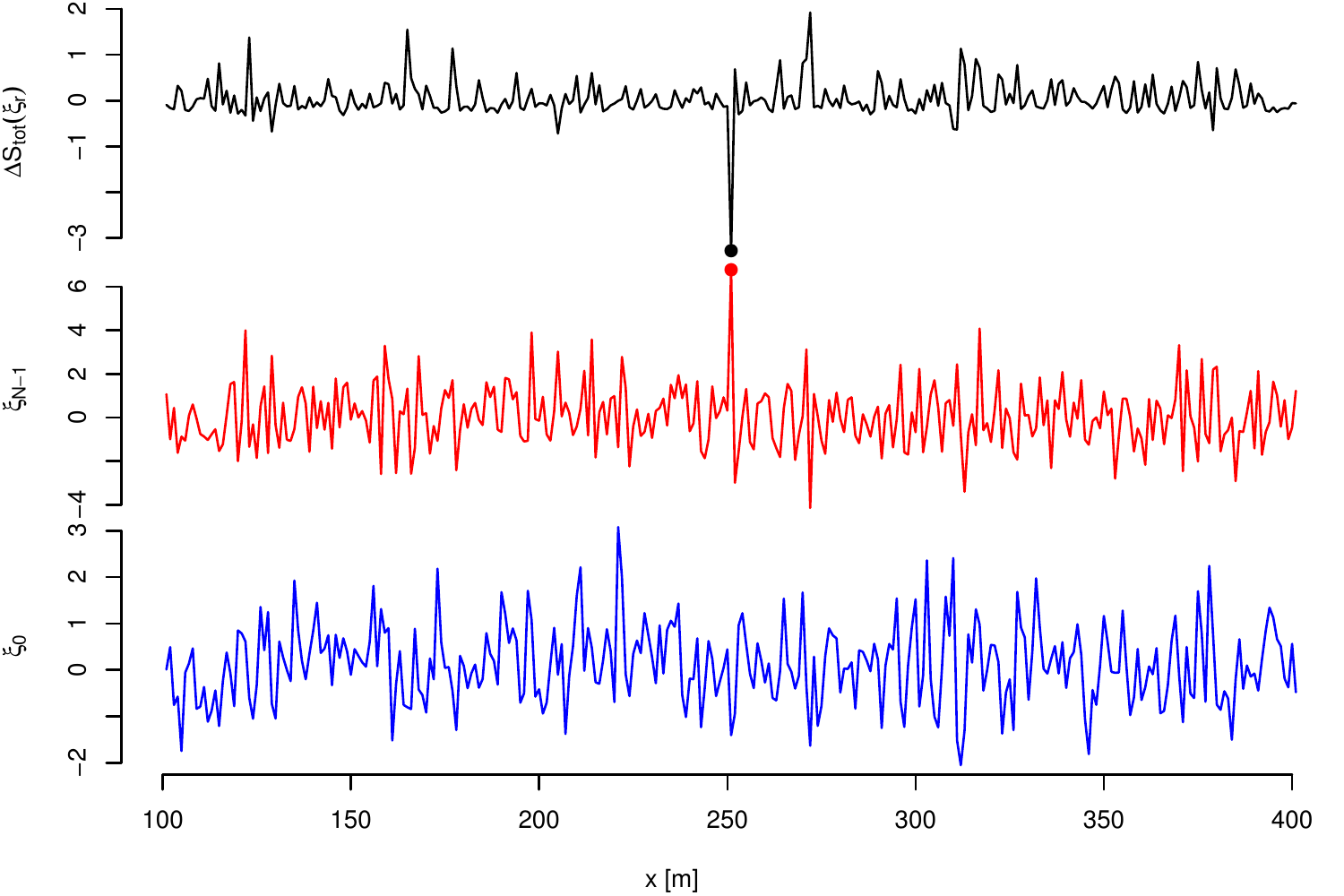}
  \caption{Time series of increments calculated from height values of
    sea waves together with the corresponding local values of the
    entropy production \cite{Hadjihoseini2017}. Note for each
    trajectory of increments one value of the entropy production is
    obtained. By solid circles the trajectory leading to the highest
    height increment on small scales is marked and related to the
    corresponding large negative entropy value. The location on the
    $x$-axis is given in units of meters, obtained by the use of Taylor's hypothesis
    with an assumed velocity of $1$ m/s. On the $y$-axis the increments for smallest scales
   are denoted by $\xi_{N-1}$  and the
    increments on large scale by $\xi_{0}$. }
  \label{Fig_EntropySeries}
\end{figure}

In particular, the concept of stochastic thermodynamics is applied to turbulent flows \cite{Nickelsen2013, Reinke2015b, Reinke2017} and sea waves \cite{Hadjihoseini2017}. The novelty here is that concepts of non-equilibrium thermodynamics known to hold for microscopic systems are shown to be valid also for such macroscopic systems.  These concepts enable to determine an entropy production of the cascade process. In particular for every individual  trajectory $\xi(\cdot)={\xi_r; r=r_0,...,r_N}$ of the increments evolving from large to small scales, a total entropy production $\Delta S_{tot}$ can be defined  by
\begin{eqnarray}
\Delta S_{tot}\left[u(\cdot) \right]  
&=&\Delta S_{med} +\Delta S_{sys} \\
&=& - \int_{r_0}^{r_N} \partial_r \xi_r \partial_{\xi}\varphi(\xi_r) \;dr - ln \frac{p(\xi_{r_N}, r_N)}{p(\xi_{r_0}, r_0)}.
\label{eq:total_S}
\end{eqnarray}
The total entropy production is given by the sum of two contributions, $\Delta S_{med}$ being the entropy variation due to the surrounding medium, which depends on the evolution of $\xi(\cdot)$ through the hierarchy of length scales $r$ in the cascade.
Here $\Delta S_{sys}$ is the entropy change of the system itself.
In Equation (\ref{eq:total_S}) $\varphi(\xi_r)$ is the potential, which can be obtained from the stationary solution of the estimated Fokker-Planck equation
\begin{equation}
	\label{eq.potential}
	\varphi(\xi_r) = ln D^{(2)}\left(\xi_r,r\right) - \int_{-\infty}^{\xi_r} \frac{D^{(1)}(\xi',r)}{D^{(2)}(\xi',r)} d\xi'.
\end{equation}
Dealing with this thermodynamics (see also  \cite{Nickelsen2013,Seifert2012}) one may interpret $\partial_{\xi}\varphi(\xi(r))$ in a less formal way as a force of the medium given by the "mean field" quantities  $D^{(1)}$ and  $D^{(2)}$.
The interaction of this force on the path "velocity" $\partial_r \xi_r$ leads then to the entropy term $\Delta S_{med}$ which represents an analogue of the work done by the medium on the single event $\xi[\cdot]$, which leads to a heat exchange with the bath (for more details of this analogy see \cite{Sekimoto1998}).  The second entropy term  $\Delta S_{sys}$  may be considered as an intrinsic contribution of the trajectory.  The main point is that to each increment trajectory $\xi[\cdot]$ a value of $\Delta S_{tot}$ can be determined like shown in Fig. \ref{Fig_EntropySeries}. Thus  "microscopic" entropy fluctuations can be determined, which may show positive and negative values. Most interestingly the negative entropy events seem to be related to extreme events in the increment statistics on the smallest scales, as can be seen in Fig. \ref{Fig_EntropySeries} and as reported in \cite{Nickelsen2013,Hadjihoseini2017}. This again points in the direction that salient structures of a complex system can be a proper part of the multi-point statistics, somehow unifying the approach to complex systems by coherent structures or by statistical methods.


\begin{textbox}[h]
\section{THERMODYNAMICAL INTERPRETATION OF TURBULENCE  }

Entropy values for cascade trajectories $\xi(\cdot)$ allow a thermodynamical interpretation. 
The potential $\varphi(\xi_r) $ of equation \ref{eq:total_S} given by the drift and diffusion term $D^{(1)}(\xi)$  and $D^{(2)}(\xi)$ 
can be considered as the coupling of the trajectory or the subsystem to the bath, whereas the values of $\xi_r$ and its probabilities are the intrinsic features of the individual trajectory. The connection with a possible thermodynamical interpretation becomes
more clear if not only the velocity increments are considered but also the transferred 
 energy $\epsilon_r$ of the cascade. As shown in \cite{Renner2002b} the Langevin equation of the cascade , equ. \ref{eq:langevin} changes to 
 \begin{eqnarray}
    - r \frac{\partial}{\partial r} \, \xi_r \; & = & \;  
   - \gamma\, \xi_r  \; + \; m \, \sqrt{\epsilon_{r}}
    \; \eta(r) , \nonumber \\
    - r\frac{\partial}{\partial r} \, \epsilon_{r} \; & \propto  & \; + \,
    G \, \epsilon_{r} \; + .... \, ,
    \label{eq:TwoDimLangevinResult}
\end{eqnarray}
 where $\gamma, G$ and $m$ are positive vales which may depend on $r$. Note that now the increment process becomes 
 purely additive, a well known effect \cite{Gagne94,Naert97}. Such an in-stationary ($r$-dependent) Langevin equation can be interpreted in a thermodynamic way, following \cite{Sekimoto1998}. Most interestingly $\epsilon_{r}$ corresponds to the temperature. As $\epsilon_{r}$ devolves its own fluctuations, the cascades can be considered as a mixture of temperatures. Note that after equation \ref{eq:TwoDimLangevinResult} these energy or temperature fluctuations increase in the cascade evolution to smaller scales. In this way the cascade pictures of Kolmogorov \cite{Kolmogorov1962} and Castaing \cite{Castaing1990} are set in a new thermodynamic context.
 
\end{textbox}
	
\begin{figure}
  \includegraphics[width=0.4\linewidth]{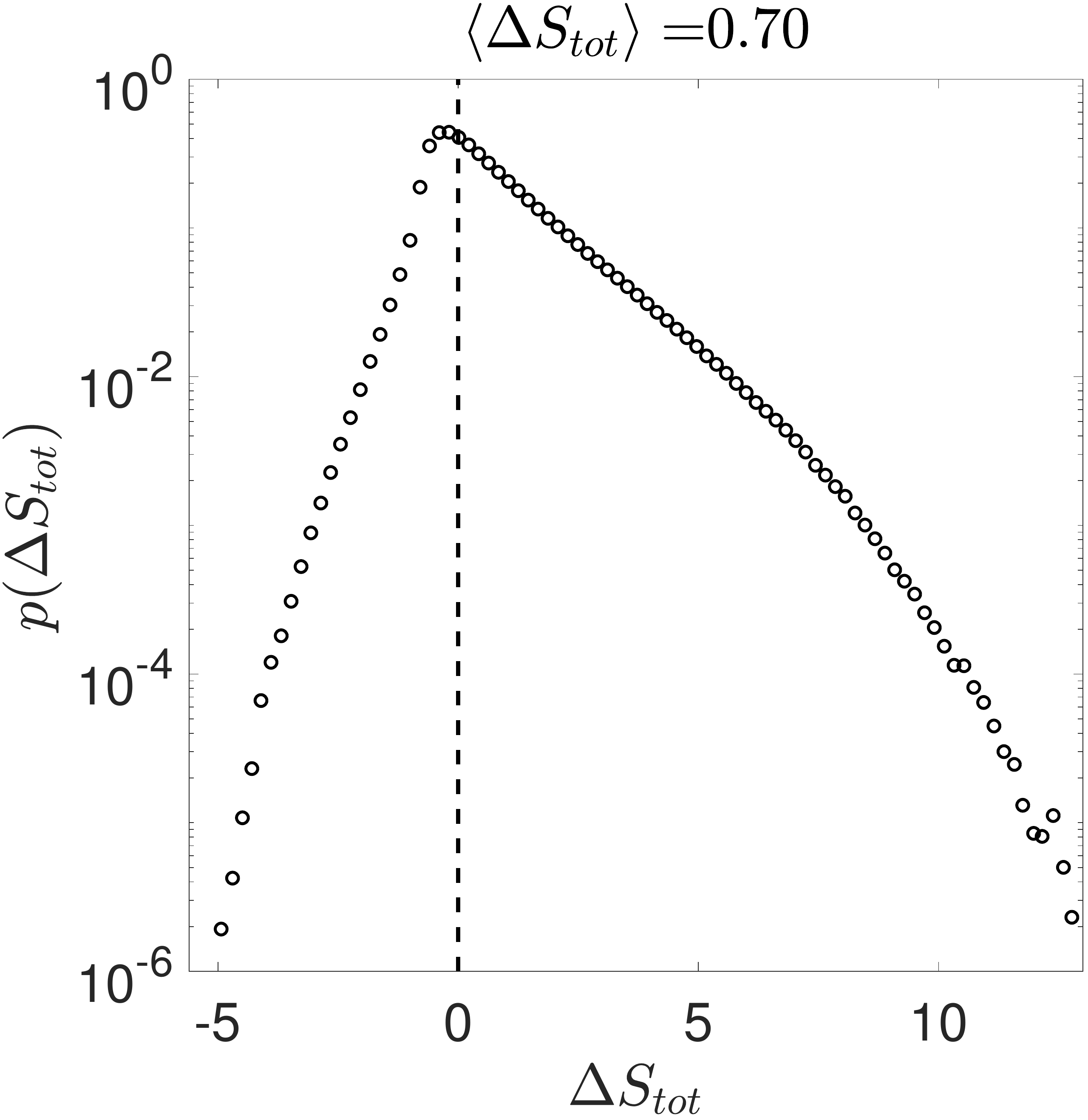}
    \includegraphics[width=.4\linewidth]{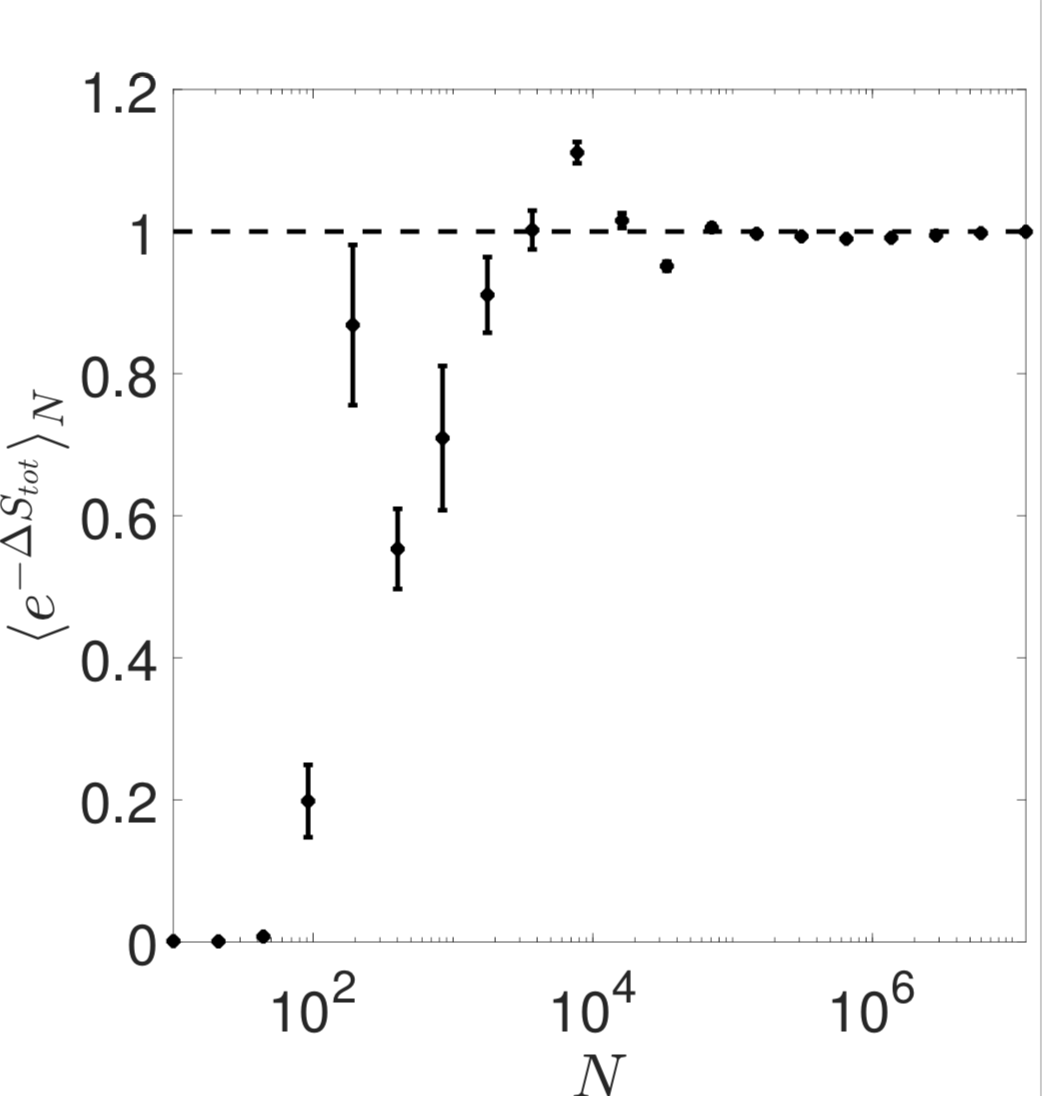}

  \caption{
  (a) Probability density function of the total entropy
    production $\Delta S_\mathit{tot}$ obtained from turbulent data,
   after \cite{Fuchs2018}. The mean of the distribution is
    positive, which means that in average the entropy increases,
    in accordance with the second law of thermodynamics. For our system here
    we find quite big fluctuations of the entropy values and  
    pronounced probabilities for negative entropy events exist. Such a
    distribution is a typical result for data like shown in
   Fig.~\ref{Fig_EntropySeries}.   
    (b) Convergence of the exponential of the entropy production
    $\langle e^{-\Delta S_{tot}} \rangle$ for turbulence data,
    following the Integral Fluctuation Theorem (IFT) in
    Eq.~\ref{eq:IFT}. Shown is the evolution of
    $\langle e^{-\Delta S_{tot}} \rangle_{N} $ as a function of the
    number $N$ of trajectories $\xi(\cdot)$ and its convergence to the
    value 1. This result shows that the IFT is fulfilled within an
    accuracy of 1$\%$ and better, see \cite{Fuchs2018}.    
    Most interestingly the IFT implies that the probabilities of the 
    negative and positive entropy events are not arbitrary, but balanced by the
    IFT, which puts much weight on the negative events.
    In other words, events with negative entropy values must be accompanied by many events with
    positive entropy to fulfil the IFT law.   
     }
  \label{Fig_EntropyPDF}
\end{figure}

If the complex structure is described correctly by the Fokker-Planck equation, the statistics of the entropy values should fulfill the integral fluctuation theorem (IFT)
\begin{equation}
\langle e^{-\Delta S_{tot}} \rangle_{N} = 1 ,
\label{eq:IFT}
\end{equation}
a fundamental entropy law of non-equilibrium thermodynamics, cf. \cite{Seifert2012}.
 Here $\langle  \cdots \rangle_N$ denotes the average over many different trajectories for the increments.
 In Fig. \ref{Fig_EntropyPDF} the distribution of the entropy production values for an experimental data set of a turbulent flow \cite{Renner2001} is shown. Clearly the mentioned positive and negative entropy values can be seen. The mean value of this distribution $\langle \Delta S_{tot} \rangle$ is positive. By a weighting function $e^{-\Delta S_{tot}} $ negative entropy values contribute much more and must be compensated by many large positive $\Delta S_{tot}$ values so that the IFT Equation (\ref{eq:IFT}) is fulfilled. 
Thus the IFT is a relation which expresses the balance between the relative frequency of entropy-consuming ($\Delta S_{tot} <0$) and entropy-producing ($\Delta S_{tot} >0$) trajectories associated with the stochastic evolution of increment trajectories $\xi(\cdot)$ (individual stochastic trajectories).

%

 In Fig. \ref{Fig_EntropyPDF}b the obtained values from Equation (\ref{eq:IFT}) for increasing numbers of trajectories are shown.
The convergence to the absolute value of 1 is well obtained. Already after several thousand values the IFT gets fulfilled.

Another interesting result is obtained for sea waves \cite{Hadjihoseini2017}. It has been found that extreme events, namely the rogue waves, are characterised by negative entropy values. Comparing different states of the sea waves it could be shown that the statistics $W(\Delta S_{tot})$ change significantly from one state to the other, although for both cases the IFT was fulfilled in high quality. Coming back to the point that the IFT somehow balances the negative and positive entropy events. As the negative entropy events are correlated to large waves, like the rogue waves, we see that this non-equilibrium thermodynamics together with the stochastic cascade process grasps both, the statistics and the localised structure of the complex disordered system.



\section{CONCLUSIONS}
\label{Sec:Conc}

The leading topic of this work was the characterisation of complex systems and the question if an understanding of the complexity can 
be achieved by structures as basic elements or if higher order statistics are needed. For two examples, namely turbulence and sea waves, we showed how these two aspects of structure and statistics are interwoven. The description of the multi-point statistics by a stochastic process of a cascade, or, respectively, by a Fokker-Planck equation evolving in scale, allows to generate surrogate data sets as well as to determine entropy values for all data points. Most interestingly for our considered macro-systems large fluctuations of these entropy values are found, as it is up to now mainly discussed for mirco-systems cf. \cite{Seifert2012}. 
We see how this concept of micro-systems can fruitfully be applied to our considered macro-systems. A key element is the Markov property in scale and the corresponding derivation of the Fokker-Planck equation. It is the Fokker-Planck equation which leads to  a general law of  non-equilibrium thermodynamics, namely, the integral fluctuation theorem, which is fulfilled for our data with an accuracy of 1 $\%$ and better. On the one hand the validity of this integral fluctuation theorem can be taken as an evidence of the consistency of our whole approach. On the other hand the integral fluctuation theorem expresses mathematically the balance between negative and positive entropy values. Each negative entropy values must be compensated by many positive entropy values to fulfil the integral fluctuation theorem. The exponential weight of the theorem means that each negative entropy values must be compensated by many positive entropy values. This seems to be also true for the structures of the complex systems, as we showed that the negative entropy values are connected with the large small-scale structures, which are the challenging properties for turbulence and waves.
Thus we conclude that our work presents a new consistent approach to the mutuality of order and stochasticity in complex systems.

\newpage

\begin{summary}[SUMMARY POINTS]
\begin{enumerate}
\item N-point statistics representing an all-encompassing probabilistic approach to complex systems can be expressed by N-1 scale increment statistics. The increment statistics allows a hierarchical ordering. If the increment statistics only depend on increments of the neighbouring scale,  a three point closure of the N-point statistics is achieved.
\item The three point closure of the N-point statistics is equivalent to a stochastic process in scale with Markov property, for which the process equations can be estimated via Kramers-Moyal coefficients from empirical or measured data. If Langevin noise is present a non-stationary Fokker-Planck equation for the cascade process in scale is obtained.
\item The knowledge of the non-stationary scale-dependent Fokker-Planck equation allows to generate numerically new data sets with the same N-point statistics and to forecast single events, most interestingly also extreme events. 
\item Based on the non-stationary scale-dependent Fokker-Planck equation the entropy production of the cascade trajectory can be defined, for which the rigorous integral fluctuation theorem holds. Thus a connection with non-equilibrium thermodynamics is given, which balances the occurrence of negative and positive entropy events. 
\end{enumerate}
\end{summary}

\begin{issues}[FUTURE ISSUES]
\begin{enumerate}
\item The non-stationary scale-dependent  stochastic description of N-point statistics can be generalised in a straight forward way to higher dimensional quantities $\vec{q}$, like complex turbulent velocity fields, see \cite{siefert2006}. Here the Fokker-Planck equation depends on different
variables of the vector field and $D^{(1)}$ becomes a vector, $D^{(2)}$ is a diffusion matrix. The problem how to extend this approach to  two- or three dimensional spaces instead of the one dimensional cut, for which a hierarchical ordering is evident remains open.
\item It is a challenge to work out a meaningful non-equilibrium thermodynamics of these complex structures, relating it to quantities like energies of the systems. It should also be noted that there have already been different tries to set up thermodynamical approaches to complex systems. A relation between those would be important.
\item Often complex systems are also described by nonlinear partial differential equations. Is there a possibility to derive the non-stationary scale-dependent stochastic process equation directly from the partial differential equations?  This would unify at least for these systems two different ways of description.
\end{enumerate}
\end{issues}


\section*{ACKNOWLEDGMENTS}
The authors acknowledge funding from the VolkswagenStiftung, the German Science Foundation DFG, and the Ministry for Science and Culture of the State of Lower Saxony (MWK) as well as helpful and inspiring discussions with Bernard Castaing,
 Christian Behnken, Andreas Engel, Jannik Ehrich, Jan and Rudolf Friedrich, Andr{\'e} Fuchs, Mathieu Gibert, Alain Girard, Hauke H\"ahne, Ali Hadjihossini, Daniel Nickelsen, and Nico Reinke. J.P. acknowledges support from the Laboratoire d’excellence LANEF in Grenoble (ANR-10-LABX-51-01). The authors devote this article to Rudolf Friedrich, with whom we had the pleasure to work out  many aspects of this works.

%
\section*{LITERATURE\ CITED}

%
%
%

\bibliographystyle{ar-style4}
\bibliography{LitV11}

\begin{thebibliography}{70}
\expandafter\ifx\csname natexlab\endcsname\relax\def\natexlab#1{#1}\fi

\bibitem{Argyris2015}
Argyris J, Faust G, Haase M, Friedrich R. 2015.
An exploration of dynamical systems and chaos.
New York: Springer

\bibitem{heslot1987transitions}
Heslot F, Castaing B, Libchaber A. 1987.
\textit{Physical Review A} 36:5870

\bibitem{Haken1}
Haken H. 2004.
Synergetics, introduction and advanced topics.
Heidelberg, New York: Springer

\bibitem{bar1997dynamics}
Bar-Yam Y. 1997.
Dynamics of complex systems.
vol. 213.
Addison-Wesley Reading, MA

\bibitem{Friedrich2011}
Friedrich R, Peinke J, Sahimi M, Tabar MRR. 2011.
\textit{Phys. Report} 506:87--162

\bibitem{Frisch2001}
Frisch U. 2001.
Turbulence: the legacy of a. n. kolmogorov.
Cambridge University Press

\bibitem{Davidson2004}
Davidson PA. 2004.
{Turbulence: an introduction for scientists and engineers}.
Oxford University Press, USA

\bibitem{Pope2000}
Pope SB. 2000.
Turbulent flows.
Cambridge University Press

\bibitem{Clay}
Clay. 2000.
Millennium problems.
Clay Mathematics Institute: http://www.claymath.org/millennium-problems

\bibitem{Nazarenko2016}
Nazarenko S, Lukaschuk S. 2016.
\textit{Annual Review of Condensed Matter Physics} 7:61--88

\bibitem{Onorato2013}
Onorato M, Residori S, Bortolozzo U, Montina A, Arecchi T. 2013 528:47–89

\bibitem{Akhmediev2009}
Akhmediev N, Ankiewicz A, Taki M. 2009.
\textit{Physics Letters A} 373:675--678

\bibitem{Kantz2004}
Kantz H, Schreiber T. 2004.
Nonlinear time series analysis.
Cambridge nonlinear science series. Cambridge University Press

\bibitem{Friedrich1997}
Friedrich R, Peinke J. 1997{\natexlab{a}}.
\textit{Phys.~Rev.~Lett.} 78:863

\bibitem{Peinke1996}
Peinke J, Friedrich R, Chill{\`a} F, Chabaud B, Naert A. 1996.
\textit{Zeitschrift f{\"u}r Physik B Condensed Matter} 101:157--159

\bibitem{Friedrich1997b}
Friedrich R, Peinke J. 1997{\natexlab{b}}.
\textit{Physica D} 102:147

\bibitem{amblard1999cascade}
Amblard PO, Brossier JM. 1999.
\textit{The European Physical Journal B-Condensed Matter and Complex Systems}
  12:579--582

\bibitem{Castaing1990}
Castaing B, Gagne Y, Hopfinger E. 1990.
\textit{Physica D: Nonlinear Phenomena} 46:177 -- 200

\bibitem{Friedrich2012}
Friedrich R, Peinke J, Reza Rahimi~Tabar M. 2012.
Fluctuations, importance of: Complexity in the view of stochastic processes.
New York, NY: Springer New York,  1131--1154

\bibitem{Nawroth2006}
Nawroth A, Peinke J. 2006.
\textit{Physics Letters A} 360:234 -- 237

\bibitem{Nawroth2010}
Nawroth AP, Friedrich R, Peinke J. 2010.
\textit{New Journal of Physics} 12:083021

\bibitem{Stresing2010}
Stresing R, Peinke J. 2010.
\textit{New Journal of Physics} 12:103046

\bibitem{Taylor1938}
Taylor GI. 1938.
\textit{Proceedings of the Royal Society of London A: Mathematical, Physical
  and Engineering Sciences} 164:476--490

\bibitem{Waechter2004b}
Waechter M, Kouzmitchev A, Peinke J. 2004.
\textit{Physical Review E}

\bibitem{Waechter2004}
Waechter M, Riess F, Schimmel T, Wendt U, Peinke J. 2004.
\textit{The European Physical Journal~B} 41:259

\bibitem{FriedrichJDiss}
Friedrich J. 2017.
\textit{PhD-Dissertation, Ruhr-Universit\"at Bochum}

\bibitem{Renner2001}
Renner C, Peinke J, Friedrich R. 2001{\natexlab{a}}.
\textit{J.~Fluid Mech.} 433:383--409

\bibitem{Friedrich1998}
Friedrich R, Zeller J, Peinke J. 1998.
\textit{Europhysics Letters} 41:153

\bibitem{Tutkun2004}
Tutkun M, Mydlarski L. 2004.
\textit{New Journal of Physics} 6:49

\bibitem{Lueck2006}
{L{\"u}ck} S, Renner C, Peinke J, Friedrich R. 2006.
\textit{Physics Letters A} 359:335--338

\bibitem{Marcq2001}
Marcq P, Naert A. 2001.
\textit{Physics of Fluids} 13:2590 -- 2595

\bibitem{Hadjihosseini2016}
Hadjihosseini A, Wächter M, Hoffmann NP, Peinke J. 2016.
\textit{New Journal of Physics} 18:013017

\bibitem{muzy1993multifractal}
Muzy JF, Bacry E, Arneodo A. 1993.
\textit{Physical review E} 47:875

\bibitem{farge2006wavelets}
Farge M, Schneider K. 2006.
\textit{Encyclopedia of Mathematical Physics} :408--420

\bibitem{lovejoy2013weather}
Lovejoy S, Schertzer D. 2013.
The weather and climate: emergent laws and multifractal cascades.
Cambridge University Press

\bibitem{Risken1984}
Risken H. 1984.
The fokker-planck equation.
Heidelberg: Springer

\bibitem{hanggi1982}
H{\"a}nggi P, Thomas H. 1982.
\textit{Physics Reports} 88:207--319

\bibitem{Gardiner1998}
Gardiner C. 1998.
Thandbook of stochastic methods for physics, chemistry and the natural
  sciences.
Berlin: Springer

\bibitem{RennerDiss}
Renner C. 2002.
\textit{PhD-Dissertation, Universit\"at Oldenburg}

\bibitem{Renner2001finance}
Renner C, Peinke J, Friedrich R. 2001{\natexlab{b}}.
\textit{Physica A: Statistical Mechanics and its Applications} 298:499 -- 520

\bibitem{Einstein1905}
Einstein A. 1905.
\textit{Annalen der Physik} 322:549--560

\bibitem{Dubrulle2000}
Dubrulle B. 2000.
\textit{Eur. Phys. J. B} 14:757–771

\bibitem{Kolmogorov1931}
Kolmogorov AN. 1931.
\textit{Math. Ann.} 104:415--458

\bibitem{Reza2000}
Davoudi J, Tabar MRR. 2000.
\textit{Physical Review E} 61:6563

\bibitem{Gottschall2008}
Gottschall J, Peinke J. 2008.
\textit{New Journal of Physics} 10:083034

\bibitem{Honisch2011}
Honisch C, Friedrich R. 2011.
\textit{Phys. Rev. E} 83:066701

\bibitem{siefert2003quantitative}
Siefert M, Kittel A, Friedrich R, Peinke J. 2003.
\textit{EPL (Europhysics Letters)} 61:466

\bibitem{bottcher2006reconstruction}
B{\"o}ttcher F, Peinke J, Kleinhans D, Friedrich R, Lind PG, Haase M. 2006.
\textit{Physical review letters} 97:090603

\bibitem{lehle2011analysis}
Lehle B. 2011.
\textit{Physical Review E} 83:021113

\bibitem{lehle2018analyzing}
Lehle B, Peinke J. 2018.
\textit{Physical Review E} 97:012113

\bibitem{Nawroth2007}
{Nawroth} AP, {Peinke} J, {Kleinhans} D, {Friedrich} R. 2007.
\textit{Phys. Rev. E.} 76

\bibitem{Anvari2016}
Anvari M, Tabar MRR, Peinke J, Lehnertz K. 2016.
\textit{Sci. Rep.} 6:35435

\bibitem{Castaing1996}
Castaing B. 1996.
\textit{J. Phys. II France} 6:105--114

\bibitem{Friedrich1997c}
Friedrich R, Peinke J, Naert A. 1997.
\textit{Z. Naturforsch} 52 a:588 -- 592

\bibitem{Kolmogorov1962}
Kolmogorov AN. 1962.
\textit{Journal of Fluid Mechanics} 13:82--85

\bibitem{NickelsenDiss}
Nickelsen D. 2014.
\textit{PhD-Dissertation, Universit\"at Oldenburg}

\bibitem{Hallerberg2007}
Hallerberg S, Altmann EG, Holstein D, Kantz H. 2007.
\textit{Phys. Rev. E} 75:016706

\bibitem{Hadjihoseini2017}
Hadjihoseini A, Lind PG, Mori N, Hoffmann NP, Peinke J. 2017.
\textit{EPL (Europhysics Letters)} 120:30008

\bibitem{Laval2001}
Laval JP, Dubrulle B, Nazarenko S. 2001.
\textit{Physics of Fluids} 13:1995--2012

\bibitem{brown1982information}
Brown TM. 1982.
\textit{Journal of Physics A: Mathematical and General} 15:2285

\bibitem{Seifert2012}
Seifert U. 2012.
\textit{Reports on Progress in Physics} 75:126001

\bibitem{Nickelsen2013}
Nickelsen D, Engel A. 2013.
\textit{Physical Review Letters} 110:214501

\bibitem{Reinke2015b}
Reinke N, Nickelsen D, Engel A, Peinke J. 2016.
\textit{Progress in Turbulence VI, Proceedings of the iTi Conference on
  Turbulence 2014, Springer Proceedings in Physics} 165:19 -- 25

\bibitem{Reinke2017}
Reinke N, Fuchs A, Nickelsen D, Peinke J. 2018.
\textit{Journal of Fluid Mechanics} 848:117--153

\bibitem{Sekimoto1998}
Sekimoto K. 1998.
\textit{Progress of Theoretical Physics Supplement} 130:17--27

\bibitem{Renner2002b}
Renner C, Peinke J, Friedrich R. 2002.
\textit{arXiv} :18

\bibitem{Gagne94}
Gagne Y, Marchand M, Castaing B. 1994.
\textit{Journal de Physique II} 4:1--8

\bibitem{Naert97}
Naert A, Friedrich R, Peinke J. 1997.
\textit{Physical Review E} 56:6719--6722

\bibitem{Fuchs2018}
Fuchs A, Reinke N, Nickelsen D, Peinke J. 2018.
A rigorous entropy law for the turbulent cascade. In \textit{Proceedings of
  EUROMECH-ERCOFTAC Colloquium 589: Turbulent Cascades II}, ed. M~Gorokhovski.
  Berlin, New York: Springer

\bibitem{siefert2006}
Siefert M, Peinke J. 2006.
\textit{Journal of Turbulence} 7:N50

\end{thebibliography}

%
%
%
%
%
%
%
%
%
%
%
%
%
%
%
%

\end{document}